\documentclass[11pt]{article}

\usepackage{tikz}
\usepackage{dsfont}
\usepackage{amssymb}
\usepackage{float}
\usepackage{chet}

\newcommand{\CH}{\mathcal{H}}
\newcommand{\CO}{\mathcal{O}}
\newcommand{\CG}{\mathcal{G}}
\newcommand{\CF}{\mathcal{F}}

\newcommand{\CN}{\mathcal{N}}
\newcommand{\CR}{\mathcal{R}}
\newcommand{\CJ}{\mathcal{J}}
\newcommand{\CS}{\mathcal{S}}
\newcommand{\COb}{{\bar{\mathcal{O}}}}

\newcommand{\COind}[1]{\CO_{#1_1\hspace{-0.8pt}\ldots#1_\ell;\lsp
  \dot{#1}_1\hspace{-0.8pt}\ldots\dot{#1}_\ell}}

\newcommand{\Qb}{{\bar{Q}}}
\newcommand{\qb}{{\bar{q}}}
\newcommand{\zb}{{\bar{z}}}
\newcommand{\jb}{{\bar{\jmath}}\hspace{0.9pt}}
\newcommand{\thetab}{{\bar{\theta}}}
\newcommand{\Thetab}{{\bar{\Theta}}}
\newcommand{\etab}{{\bar{\eta}}}
\newcommand{\phib}{{\bar{\phi}}}
\newcommand{\Phib}{{\bar{\Phi}}}
\newcommand{\Sb}{{\bar{S}}}
\newcommand{\alphad}{{\dot{\alpha}}}
\newcommand{\betad}{{\dot{\beta}}}
\newcommand{\xup}{{\text{x}}}
\newcommand{\vev}[1]{{\langle #1\rangle}}

\newcommand{\lsp}{\hspace{1pt}}
\newcommand{\lnsp}{\hspace{-1pt}}
\newcommand{\dtwo}{d^{\hspace{0.5pt}2}\hspace{-0.5pt}}

\date{February 2017}

\preprint{CERN-TH-2017-024}

\title{Bootstrapping Mixed Correlators\\\vspace{6pt}
in 4D $\titlemath{\CN=1}$ SCFTs}

\author{Daliang Li,$^{\!a}$ David Meltzer,$^{\!b}$ and Andreas
Stergiou$^{b,c}$}

\affiliation{$^{a}$Department of Physics and Astronomy, Johns Hopkins
University,\\\vspace{-3pt}
Charles Street, Baltimore, MD 21218, USA\\
$^{b}$Department of Physics, Yale University, New Haven, CT 06520, USA\\
$^{c}$Theoretical Physics Department, CERN, Geneva, Switzerland}

\abstract{The numerical conformal bootstrap is used to study mixed
correlators in $\mathcal{N}=1$ superconformal field theories (SCFTs) in
$d=4$ spacetime dimensions. Systems of four-point functions involving
scalar chiral and real operators are analyzed, including the case where the
scalar real operator is the zero component of a global conserved current
multiplet. New results on superconformal blocks as well as universal
constraints on the space of 4D $\mathcal{N}=1$ SCFTs with chiral operators
are presented. At the level of precision used, the conditions under which
the putative ``minimal'' 4D $\mathcal{N}=1$ SCFT may be isolated into a
disconnected allowed region remain elusive.  Nevertheless, new features of
the bounds are found that provide further evidence for the presence of a
special solution to crossing symmetry corresponding to the ``minimal'' 4D
$\mathcal{N}=1$ SCFT.}

\begin{document}

\maketitle

\toc

\newsec{Introduction}[intro]
The modern revival of the conformal bootstrap program
\cite{Rattazzi:2008pe} has led to remarkable progress in our understanding
of conformal field theories (CFTs) in $d>2$ spacetime dimensions. By
studying the constraints of crossing symmetry and unitarity, it is possible
to derive rigorous bounds on the scaling dimensions and operator product
expansion (OPE) coefficients of any CFT. This approach relies on very few
assumptions and can thus be used to study and discover theories without a
known Lagrangian description.

A striking result of the numerical conformal bootstrap is that the bounds
can develop kinks, or singularities, corresponding to known theories.  This
was observed in the 3D Ising~\cite{ElShowk:2012ht} and $\text{O}(N)$ vector
models~\cite{Kos:2013tga} and was correlated with the decoupling of certain
operators. This intuition was further developed in~\cite{El-Showk:2016mxr}.
With the introduction of multiple correlators and additional assumptions on
the number of relevant scalars, small regions surrounding the known
theories can be isolated from other solutions of the bootstrap equations,
i.e.\ the kinks become islands~\cite{Kos:2014bka, Kos:2015mba}.
Consequently, the known theory is essentially the unique consistent
solution of the crossing equations in a certain region in parameter space,
given certain mild assumptions.

In $d=4$ a kink was observed for $\mathcal{N}=1$ superconformal theories
(SCFTs) with a chiral scalar operator $\phi$ \cite{Poland:2010wg,
Poland:2011ey, Poland:2015mta}. More specifically, the scaling dimension
bound for the first real scalar in the $\bar{\phi}\times\phi$ OPE develops
a kink as a function of $\Delta_{\phi}$ at the same point where the lower
bound for the three-point function coefficient $c_{\phi\phi\phi^{2}}$
disappears. Similar behavior was also observed for theories in $2\leq d
\leq 4$ with four supercharges \cite{Bobev:2015jxa}. In
\cite{Poland:2015mta} it was conjectured that there is a 4D superconformal
field theory (SCFT) that saturates the bootstrap bounds at the kink,
referred to as the minimal 4D $\mathcal{N}=1$ SCFT.  Based on the position
of the kink and a corresponding local minimum in the lower bound on the
central charge, this minimal theory was predicted to have
$c_{\text{minimal}}=\frac{1}{9}$ and a chiral multiplet with scaling
dimension $\Delta_\phi=\frac{10}{7}$, which also satisfies the chiral ring
condition $\phi^2=0$. Various speculations about this minimal theory have
appeared \cite{Xie:2016hny, Buican:2016hnq}. In these proposals $\phi^2=0$
is explicitly satisfied, but the central charge and the critical
$\Delta_\phi$ have not been successfully reproduced. As a result, the
identity of this minimal theory remains elusive.

Motivated by this open problem, we study here the mixed correlator
bootstrap for 4D $\mathcal{N}=1$ theories for the system of correlators $\{
\langle \bar{\phi} \phi \bar{\phi} \phi\rangle, \langle\bar{\phi} R \phi R
\rangle, \langle RRRR\rangle\}$, where $R$ is a generic real scalar and
$\phi$ is a chiral scalar. We consider both the case where $R$ is the first
real scalar in the $\bar{\phi}\times\phi$ OPE (beyond the identity operator
of course), and that where $R$ saturates the unitarity bound. In the latter
case it sits in a linear multiplet, which we will label by $J$. The
bootstrap equations for the $ \langle \bar{\phi} \phi \bar{\phi}
\phi\rangle$ correlator were first considered in~\cite{Poland:2010wg} and
for $\langle JJJJ\rangle $ in~\cite{Berkooz:2014yda}, and for $\langle
RRRR\rangle$ in~\cite{Khandker:2014mpa}. Here we present new results for
the superconformal blocks of $\langle\bar{\phi} R \phi R \rangle$ and
$\langle\bar{\phi} J \phi J \rangle$.  To be precise, we find
superconformal blocks when the superconformal primary of the exchanged
multiplet appears in a $(j,\bar{\jmath})$ representation of
$\text{SO}(3,1)$, with $j\neq\bar{\jmath}$. In this case the corresponding
superconformal primary does not appear in the OPE of the external
operators, but some of its superconformal descendants do. We also compute
superconformal blocks of superconformal primaries in integer-spin
representations; our results agree with the
literature~\cite{Khandker:2014mpa, Fitzpatrick:2014oza, Li:2016chh}.

Our main results are new numerical constraints on 4D $\mathcal{N}=1$
theories.  Studying the single correlator $\langle JJJJ\rangle $, where $J$
corresponds to a $\text{U}(1)$ linear multiplet, we improve upper bounds on
the OPE coefficients for $\langle JJJ\rangle$ and $\langle JJV\rangle$
where $V$ is the spin-one multiplet containing the stress-energy tensor
$T^{\mu\nu}$. We also study these bounds as a function of the first
unprotected scalar in the $J\times J$ OPE, deriving an upper bound on this
operators scaling dimension and the $\langle JJ O\rangle$ OPE coefficient.
With the mixed correlator system for $\phi$ and $R$, with $R$ the first
real scalar in the $\bar{\phi}\times\phi$ OPE, we will derive stronger
lower bounds on the central charge $c$ and upper and lower bounds on
$c_{\phi \bar{\phi} R}$. In both cases we find interesting features near
the minimal $\mathcal{N}=1$ point.  Finally, studying the mixed correlator
system for $\phi$ and $J$ we will derive new bounds on $c_{\phi
\bar{\phi}J}$ and $c_{\bar{\phi}J(\phi J)}$ where $(\phi J)$ is the second
scalar appearing in the $\phi \times J$ OPE.

In sections \ref{secSCB} and \ref{secJBl} we give the complete set of
conformal blocks for the mixed correlator system involving a generic real
scalar multiplet $R$ and the linear multiplet $J$ respectively. In sections
\ref{crossRel} and \ref{crossRelJ} we give the corresponding crossing
relations for $R$ and $J$. In section \ref{BoundR} we present results for
the $\phi$ and $R$ system. In section \ref{BoundJ} we present results for
the $\phi$ and $J$ system. In appendix \ref{appApprox} we will go over the
approximations used in the numerical implementation of the crossing
equations and in appendix \ref{appSBlocks} we give some details on the
derivation of the superconformal blocks.

\newsec{Four-point functions, conformal and superconformal blocks}[secSCB]
In this section we present our results for the superconformal block
decomposition of the various four-point functions used in our bootstrap
analysis. In particular we include results for the four-point function
$\vev{\phib(x_1)\lsp\phi(x_2)\lsp \phib(x_3)\lsp \phi(x_4)}$, first
obtained in \cite{Poland:2010wg, Vichi:2011ux}, and new results for the
four-point function $\vev{\phib(x_1)\lsp R(x_2)\lsp \phi(x_3)\lsp R(x_4)}$,
with $R$ a real operator, in the $\phib\times R$ channel. In our numerical
analysis we also use the four-point function $\vev{\phib(x_1)\lsp
R(x_2)\lsp \phi(x_3)\lsp R(x_4)}$ in the $\phib\times\phi$ channel, results
for which were first obtained in~\cite{Khandker:2014mpa} (see
also~\cite{Li:2016chh}).  This forces us to also consider $\vev{R(x_1)\lsp
R(x_2)\lsp R(x_3)\lsp R(x_4)}$, where again we use results
of~\cite{Khandker:2014mpa}.

Four-point functions can be reduced and computed via the OPE.  Consider the
four-point function $\vev{\CO_i(x_1)\lsp \CO_j(x_2)\lsp \CO_k(x_3)\lsp
\CO_l(x_4)}$ where all operators are conformal primary.  We can use the
OPEs $\CO_i(x_1)\times\CO _j(x_2)$ and $\CO_k(x_3)\times \CO_l(x_4)$ to
obtain
\eqna{\vev{\CO_i(x_1)\lsp \CO_j(x_2)\lsp \CO_k(x_3)\lsp\CO_l(x_4)}=
\frac{1}{r_{12}^{\,\Delta_i+\Delta_j}\,r_{34}^{\,\Delta_k+\Delta_l}}
&\left(\frac{r_{24}}{r_{14}}\right)^{\Delta_{ij}}
\left(\frac{r_{14}}{r_{13}}\right)^{\Delta_{kl}}\\
&\times\sum_{\substack{\text{conformal}\\\text{primaries}\\\CO_m}}
\delta_{mn}c_{ij}{\lnsp}^{m}c_{kl}{\lnsp}^n \lsp
g_{\Delta_m,\lsp\ell_m}^{\Delta_{ij},\lsp\Delta_{kl}}(u,v)\,, }[FourPoint]
where $r_{ij}=(x_{ij}^{\,2})^{\frac12}$, $x_{ij}=x_i-x_j$,
$\Delta_{ij}=\Delta_i-\Delta_j$ and similarly for $\Delta_{kl}$, $\Delta_m,
\ell_m$ is the scaling dimension and spin of the exchanged operator, and
\eqn{u=\frac{x_{12}^{\,2}\lsp x_{34}^{\,2}}{x_{13}^{\,2}\lsp
x_{24}^{\,2}}=z\zb\,,\qquad v=\frac{x_{14}^{\,2}\lsp
x_{23}^{\,2}}{x_{13}^{\,2}\lsp x_{24}^{\,2}}=(1-z)(1-\zb)}[ConfXRatios]
are the two independent conformally-invariant cross ratios constructed out
of four points in space.  The conformal blocks $g_{\Delta,\lsp
\ell}^{\Delta_{ij},\lsp\Delta_{kl}}$ are functions that account for the sum
over conformal descendants.  They are given by \cite{Dolan:2000ut,
Dolan:2003hv}\foot{Compared to their original definition we drop a factor
of $2^{-\beta}$ in $g_{\alpha,\lsp\beta}^{\gamma,\lsp\delta}$, i.e.\
$(g_{\alpha,\lsp\beta}^{\gamma,\lsp\delta})^{\text{here}}=2^\beta
(g_{\alpha,\lsp\beta}^{\gamma,\lsp\delta})^{\text{D\&O}}$, by rescaling
appropriately the OPE coefficients in \FourPoint.}
\eqn{\begin{gathered} g_{\alpha,\lsp\beta}^{\gamma,\lsp\delta}(z,\zb)=
(-1)^\beta\frac{z\bar z}{z-\bar z}\big(k_{\alpha+\beta}^{\gamma,\lsp
\delta}(z)\lsp k_{\alpha-\beta-2}^{\gamma,\lsp\delta}
(\zb)-(z\leftrightarrow\zb)\big)\,,\\
k_\alpha^{\beta,\gamma}(x)=x^{\alpha/2}\lsp{}_2F_{1}
\big(\tfrac12(\alpha-\beta),\tfrac12(\alpha+\gamma);\alpha;x\big)\,.
\end{gathered}}[ConfBlocks]

In $\CN=1$ superconformal theories some of the conformal primaries in the
sum in \FourPoint are superconformal descendants, and so their
contributions to the four-point function can also be accounted for by
computing ``superconformal blocks''. The dimensions of the exchanged
operators are constrained by unitarity to be~\cite{Flato:1983te,
Dobrev:1985qv}
\eqn{\Delta\geq\left|q-\qb-\tfrac12(j-\jb)\right|
+\tfrac12(j+\jb)+2\,,}[GenUniBound]
where $(\frac12j,\frac12\jb)$ is the representation of $\CO$ under the
Lorentz group, viewed here as $\text{SU}(2)\times\text{SU}(2)$, and $q$ and
$\qb$ give the scaling dimension and R-charge of an operator via
\eqn{\Delta=q+\qb\,,\qquad R=\tfrac23(q-\qb)\,.}[]

\subsec{Four-point function \texorpdfstring{$\langle\phib(x_1)\lsp\phi(x_2)
\lsp\phib(x_3)\lsp\phi(x_4)\rangle$}{<phib phi phib phi>}}
The four-point function $\vev{\phib(x_1)\lsp \phi(x_2)\lsp \phib(x_3)\lsp
\phi(x_4)}$ involving the chiral operator $\phi$ and its complex conjugate
can be expressed in terms of $12\rightarrow34$ contributions
as~\cite{Poland:2010wg}
\eqn{\vev{\phib(x_1)\lsp \phi(x_2)\lsp \phib(x_3)\lsp \phi(x_4)}
=\frac{1}{r_{12}^{\,2\Delta_\phi}\,r_{34}^{\,2\Delta_\phi}}
\sum_{\substack{\text{superconformal}\\\text{primaries}\\\CO_\ell
\in\phib\times\phi}}\hspace{-4pt}|c_{\phib\phi\CO_\ell}|^2\lsp
(-1)^\ell\lsp\CG_{\Delta,\lsp\ell}
^{\phib\phi;\lsp \phib\phi}(u,v)\,,}[chiantichi]
where we used $c_{\phib\phi\CO_\ell}=(-1)^\ell c_{\phib\phi\CO_\ell}^\ast$
and
\eqn{\CG_{\Delta,\lsp\ell}^{\phib\phi;\lsp \phib\phi}=
g_{\Delta,\lsp\ell}
-c_1\lsp g_{\Delta+1,\lsp\ell+1}
-c_2\lsp g_{\Delta+1,\lsp\ell-1}
+c_1 c_2\lsp g_{\Delta+2,\lsp\ell}\,,
\qquad g_{\alpha,\beta}\equiv g_{\alpha,\beta}^{0,0}\,,}[Gphibphi]
with
\eqn{c_1=\frac{\Delta+\ell}{4(\Delta+\ell+1)}\,,\qquad
c_2=\frac{\Delta-\ell-2}{4(\Delta-\ell-1)}\,.}[]
The unitarity bound here is $\Delta\geq\ell+2$ and, when it is saturated,
$c_2$ becomes zero.

If we flip the last two operators in the four-point function and consider
$\vev{\phib(x_1)\lsp \phi(x_2)\lsp \phi(x_3)\lsp \phib(x_4)}$, then we can
write, in the $12\rightarrow34$ channel,
\eqn{\vev{\phib(x_1)\lsp \phi(x_2)\lsp \phi(x_3)\lsp \phib(x_4)}
=\frac{1}{r_{12}^{\,2\Delta_\phi}\,r_{34}^{\,2\Delta_\phi}}
\sum_{\substack{\text{superconformal}\\\text{primaries}\\\CO_\ell
\in\phib\times\phi}}\hspace{-4pt}|c_{\phib\phi\CO_\ell}|^2\lsp
\CG_{\Delta,\lsp\ell}^{\phib\phi;\lsp \phi\phib}(u,v)\,,}[]
where we used $c_{\phi\phib\CO_\ell}=c_{\phib\phi\CO_\ell}^\ast$ and
\eqn{\CG_{\Delta,\lsp\ell}^{\phib\phi;\lsp \phi\phib}=
g_{\Delta,\lsp\ell}
+c_1\lsp g_{\Delta+1,\lsp\ell+1}
+c_2\lsp g_{\Delta+1,\lsp\ell-1}
+c_1 c_2\lsp g_{\Delta+2,\lsp\ell}\,.}[Gphiphib]
The difference between \Gphibphi and \Gphiphib is just in the sign of the
$g_{\Delta+1,\lsp\ell\pm1}$ contributions.

In this work we will also decompose $\vev{\phib(x_1)\lsp \phi(x_2)\lsp
\phi(x_3)\lsp \phib(x_4)}$ in the $14\rightarrow32$
channel~\cite{Vichi:2011ux},
\eqn{\vev{\phib(x_1)\lsp \phi(x_2)\lsp \phi(x_3)\lsp \phib(x_4)}
=\frac{1}{r_{14}^{\,2\Delta_\phi}\,r_{23}^{\,2\Delta_\phi}}
\sum_{\substack{\text{conformal}\\\text{primaries}\\\COb_\ell
\in\phib\times\phib}}\hspace{-5pt}|c_{\phib\phib\CO_\ell}|^2
\lsp\CG_{\Delta,\lsp\ell}^{\phib\phib;\lsp \phi\phi}(v,u)\,, }[chichi]
where we used $c_{\phi\phi\CO_\ell}=c_{\phib\phib\COb_\ell}^\ast$ and
\eqn{\CG_{\Delta,\lsp\ell}^{\phib\phib;\lsp
\phi\phi}=g_{\Delta,\lsp\ell}\,.}[]
In this case no superconformal block needs to be computed, but we need to
include all classes of conformal primaries that can appear in the
$\phi\times\phi$ OPE. This has been done in~\cite{Vichi:2011ux} and uses
the fact that the product $\phi\times\phi$ is chiral and that the
three-point function $\vev{\Phi(z_1)\lsp \Phi(z_2)\lsp\CO_I(z_3)}$ is
symmetric under $z_1\leftrightarrow z_2$. Here $z=(x,\theta,\thetab)$ is a
point in superspace, and the index $I$ denotes Lorentz indices. The
contributions we need to include turn out to be the superconformal primary
$\phi^2$, protected even-spin operators of the form $\bar{Q}\CO_\ell$ with
dimension $\Delta=2\Delta_\phi+\ell$, and unprotected even-spin operators
of the form $\bar{Q}^2\CO_\ell$ with dimension satisfying
$\Delta\geq|2\Delta_\phi-3|+3+\ell$. When $\Delta_\phi<\frac32$ there is a
gap in the dimensions of the unprotected and protected operators.

\subsec{Four-point function \texorpdfstring{$\langle\phib(x_1)\lsp
R(x_2)\lsp \phi(x_3)\lsp R(x_4)\rangle$}{<phib R phi R>}}
The four-point function $\vev{\phib(x_1)\lsp R(x_2)\lsp \phi(x_3)\lsp
R(x_4)}$, involving the chiral operator $\phi$ and the real operator $R$,
can be expanded in the $12\rightarrow34$ channel as
\eqn{\vev{\phib(x_1)\lsp R(x_2)\lsp \phi(x_3)\lsp R(x_4)}
=\frac{1}{r_{12}^{\,\Delta_\phi+\Delta_R}\,r_{34}^{\,\Delta_\phi+\Delta_R}}
\left(\frac{r_{24}}{r_{13}}\right)^{\Delta_\phi-\Delta_R}\hspace{-10pt}
\sum_{\COb_\ell\in\phib\times R}\hspace{-4pt}|c_{\phib R\CO_\ell}|^2
\lsp\CG_{\Delta,\lsp\ell,\lsp \Delta_\phi-\Delta_R}^{\phib R\lsp;\lsp
\phi R}(u,v)\,,}[FourPointOnes]
where $\Delta_\phi,\Delta_R$ are the scaling dimensions of $\phi,R$
respectively, $\Delta,\lsp\ell$ are the scaling dimension and spin of
$\CO$, $\bar{c}_{\phib R\CO_\ell}$ is the coefficient of the three-point
function $\vev{\phib(x_1)\lsp R(x_2)\lsp \CO_I(x_3)}$, and we use
$\bar{c}_{\phi R\COb_\ell}=\bar{c}_{\phib R\CO_\ell}^{\lsp\ast}$. As we
will see below the sum in the right-hand side of \FourPointOnes contains
contributions from multiple classes of operators.

In order to compute $\CG_{\Delta,\lsp\ell,\lsp \Delta_\phi-\Delta_R}^{\phib
R\lsp;\lsp \phi R}$ we need the general form of the three-point function
$\vev{\Phib(z_1)\lsp \CR(z_2)\lsp\CO_I(z_3)}$, where $\CO_I$ is a
superconformal primary operator. To obtain this we use the results of
\cite{Park:1997bq, Osborn:1998qu}.  To start, we note that $\Phib$ has
superconformal weights $q_\Phib=0$ and $\bar{q}_\Phib=\Delta_\phi$, while
$\CR$ has $q_\CR=\bar{q}_\CR=\frac12\Delta_\CR$. General superconformal
constraints imply that the three-point function is proportional to a
function of $X_3, \Theta_3$ and $\Thetab_3$~\cite{Osborn:1998qu},
\eqn{\vev{\Phib(z_1)\lsp \CR(z_2)\lsp\CO_I(z_3)}=
\frac{1}{x_{\bar{1}3}{\lnsp}^{2\Delta_\phi}
x_{\bar{2}3}{}^{\Delta_R}x_{\bar{3}2}{}^{\Delta_R}}\,
t_I(X_3,\Theta_3,\Thetab_3)\,,}[ThreePF]
with the homogeneity property
\eqn{\begin{gathered} t_I(\lambda\bar{\lambda}
  X,\lambda\Theta,\bar{\lambda}\Thetab)=
  \lambda^{2a}\bar{\lambda}^{2\bar{a}}\,t_I(X,\Theta,\Thetab)\,,\\
  a-2\bar{a}=\bar{q}_\Phib+\bar{q}_\CR-q_\CO\,,\qquad
  \bar{a}-2a=q_\Phib+q_\CR-\bar{q}_\CO\,.  \end{gathered}}[abara]
Quantities appearing in \ThreePF are defined as
\eqn{\begin{gathered}
  {\text{X}}_3=\frac{{\xup}_{3\bar{1}}\tilde{\xup}_{\bar{1}
2}{\xup}_{2\bar{3}}}{x_{\bar{1}3}{}^2 x_{\bar{3}2}{}^2}\,, \qquad
\xup_{\alpha\alphad}=\sigma_{\mu\alpha\alphad}x^\mu\,,\qquad
\tilde{\xup}^{\alphad\alpha}=\epsilon^{\alpha\beta}
\epsilon^{\alphad\betad}\xup_{\beta\betad}\,,\\
\Theta_3=i\left(\frac{1}{x_{\bar{1}{3}}{}^2}\xup_{3\bar{1}}\thetab_{31}
-\frac{1}{x_{\bar{2}3}{}^2}\xup_{3\bar{2}}\thetab_{32}\right),\qquad\Thetab_3=\Theta_3^\ast\,,
\end{gathered}}[Xdefn]
with $\thetab_{ij}=\thetab_i-\thetab_j$ and the supersymmetric interval
between $x_i$ and $x_j$ defined by
\eqn{x_{\bar{\imath}j}=-x_{j\bar{\imath}}\equiv x_{ij}
-i\theta_i\sigma\bar{\theta}_i
-i\theta_j\sigma\bar{\theta}_j+2i\theta_j\sigma\bar{\theta}_i\,.
}[superdistance]

Let us first assume that $\CO_I$ has $q=\frac12(\Delta+\Delta_\phi)$ and
$\bar{q}=\frac12(\Delta-\Delta_\phi)$, as would be the case if the zero
component of $\COb_I$ appeared in the $\phib\times R$ OPE. Then,
$a=\bar{a}$, which implies that $t_I$ in \ThreePF can only be a function of
the product $\Theta_3\Thetab_3$. Furthermore, the Ward identity following
from the antichirality property of $\Phib$ implies that $t_I$ cannot be a
function of $\Thetab_3$. Therefore, $t_I$ can only be a function of $X_3$
in this case.

With the constraints we just described the operator $\CO_I$ in \ThreePF is
an integer-spin traceless-symmetric superconformal primary
$\COind{\alpha}$, with the dotted and undotted indices symmetrized
independently of each other, for which we can write
\eqn{t_{\alpha_1\hspace{-0.8pt}\ldots\alpha_\ell;\lsp \alphad_1
\hspace{-0.8pt}\ldots\alphad_\ell}(X_3)=\bar{c}_{\phib R \CO_\ell}\,
\text{X}_{3\lsp(\alpha_1(\alphad_1}\!
\cdots\text{X}_{3\lsp\alpha_\ell)\alphad_\ell)}\lsp
X_3{\vphantom{X}}^{\Delta-\ell-\Delta_\phi-\Delta_R}\,,}[tI]
where the dotted indices are symmetrized independently of the undotted
ones. With \tI the $\theta$ expansion of both sides of \ThreePF can be
performed with \emph{Mathematica} by extending the code developed for the
purposes of \cite{Li:2014gpa}. We need the superconformal primary
zero-components of $\Phib$ and $\CR$, but then the possible contributions
to the three-point function come not only from the zero component of
$\COind{\alpha}$, but also from the conformal primaries in its
$\theta\thetab$ and $\theta^2\thetab^2$ components. Taking into account all
these contributions and using results of~\cite{Li:2014gpa} leads to the
superconformal block
\eqn{\bar{\CG}_{\Delta,\lsp\ell,\lsp \Delta_\phi-\Delta_R}
^{\phib R\lsp;\lsp \phi R}=
g_{\Delta,\lsp\ell}^{\Delta_\phi-\Delta_R}
+\bar{c}_1\,g_{\Delta+1,\lsp\ell+1}^{\Delta_\phi-\Delta_R}
+\bar{c}_2\,g_{\Delta+1,\lsp\ell-1}^{\Delta_\phi-\Delta_R}
+\bar{c}_1 \bar{c}_2\,g_{\Delta+2,\lsp\ell}
^{\Delta_\phi-\Delta_R}\,,\qquad
g_{\alpha,\lsp\beta}^{\gamma}\equiv g_{\alpha,\lsp\beta}^{\gamma,\lsp
\gamma}\,,}[SCBOne]
with
\eqn{\begin{gathered}
\bar{c}_1=\frac{(\Delta+\ell-\Delta_\phi)
(\Delta+\ell+\Delta_\phi-\Delta_R)^2}
{4(\Delta+\ell)(\Delta+\ell+1)(\Delta+\ell
+\Delta_\phi)}\,,\\
\bar{c}_2=\frac{(\Delta-\ell-\Delta_\phi-2)
(\Delta-\ell+\Delta_\phi-\Delta_R-2)^2}
{4(\Delta-\ell-1)(\Delta-\ell-2)(\Delta-\ell+\Delta_\phi-2)}\,.
\end{gathered}}[cOneTwo]
The unitarity bound on $\CO_\ell$ that follows from \GenUniBound is
\eqn{\Delta \ge \Delta_\phi+\ell+2\,.}[UniBound]
When the unitarity bound \UniBound is saturated, we see from \cOneTwo that
$\bar{c}_2=0$ as expected.\foot{As an aside we note here that, for a
general scalar operator $\CS$ with superconformal weights $q_\CS$ and
$\qb_\CS$, we get an expression similar to \SCBOne for the corresponding
block $\bar{\CG}_{\Delta,\lsp\ell,\lsp\Delta_\phi-\Delta_S}^{\phib
S;\lsp\phi \Sb}$, with the coefficients
\eqn{\begin{gathered}
\bar{c}_1=\frac{(\Delta+\ell-\Delta_\phi+q_\CS-\qb_\CS)
(\Delta+\ell+\Delta_\phi-q_\CS-\qb_\CS)^2}
{4(\Delta+\ell)(\Delta+\ell+1)(\Delta+\ell
+\Delta_\phi-q_\CS+\qb_\CS)}\,,\\
\bar{c}_2=\frac{(\Delta-\ell-\Delta_\phi+q_\CS-\qb_\CS-2)
(\Delta-\ell+\Delta_\phi-q_\CS-\qb_\CS-2)^2}
{4(\Delta-\ell-1)(\Delta-\ell-2)(\Delta-\ell+\Delta_\phi-q_\CS
+\qb_\CS-2)}\,.\end{gathered}}[]
}

The block $\bar{\CG}_{\Delta,\lsp\ell,\lsp \Delta_\phi-\Delta_R} ^{\phib
R\lsp;\lsp \phi R}$ we just computed constitutes merely one of the possible
contributions to the right-hand side of \FourPointOnes. Further, we note
that, in general, $R$ is an operator exchanged in the $\phib\times\phi$
OPE, and so we also need to consider the three-point function
\eqn{\langle\Phib(z_1)\lsp \CR(z_2)\lsp\Phi(z_3)\rangle=
\frac{\bar{c}_{\phib R\phi}}{x_{\bar{1}3}{\lnsp}^{2\Delta_\phi}
x_{\bar{2}3}{}^{\Delta_R}x_{\bar{3}2}{}^{\Delta_R}}
X_3{\vphantom{X}}^{-\Delta_R}\,.}[ThreePFPhi]
Since $\Phi$ has $\qb=\jb=0$, the unitarity bound \GenUniBound is modified
to $q\geq j+1$. This implies that $\Phi$ has $\Delta\geq1$. In this case we
only need to consider a conformal block $g_{\Delta_\phi,\lsp
0}^{\Delta_\phi-\Delta_R}$. Note that due to this contribution there is
always a gap in the scalar spectrum of the $\phib\times R$ OPE.

We should also consider the case where the zero component of $\COb$ does
not contribute to the $\phib\times R$ OPE. Due to the antichirality
property of $\Phib$ it is still true that there cannot be a $\Thetab_3$ in
$t_I$, but now both $\Theta_3$ and $\Theta_3^2$ are allowed.

In the first case, relevant operators are of the form
$\CO_{\alpha_1\hspace{-0.8pt}\ldots\alpha_\ell;\lsp\alphad\alphad_1
\hspace{-0.8pt}\ldots\alphad_\ell}$ for some $\ell$ and with
$q=\frac12(\Delta+\Delta_\phi-\frac32)$ and
$\qb=\frac12(\Delta-\Delta_\phi+\frac32)$, so that
$Q^\alpha\COb_{\alpha\alpha_1 \hspace{-0.8pt}\ldots \alpha_\ell;\lsp
\alphad_1 \hspace{-0.8pt}\ldots\alphad_\ell}$ is a spin-$\ell$ conformal
primary that can appear in the $\phib\times R$ OPE.\foot{The three-point
function $\vev{\Phib(z_1)\lsp R(z_2)\lsp
\CO_{\alpha_1\hspace{-0.8pt}\ldots\alpha_\ell;\lsp\alphad
\alphad_1\hspace{-0.8pt}\ldots\alphad_\ell}(z_3)}$ is proportional to
$\Theta_3$, for \abara gives $2(a-\bar{a})=1$.}  In this case
\eqn{t_{\alpha_1\hspace{-0.8pt}\ldots\alpha_\ell;\lsp\alphad \alphad_1
\hspace{-0.8pt}\ldots\alphad_\ell}(X_3)=\hat{c}_{\phib R \CO_\ell}\,
\Theta_3{\!}^\alpha\text{X}_{3\lsp\alpha(\alphad}\lsp
\text{X}_{3\lsp(\alpha_1\alphad_1}\!
\cdots\text{X}_{3\lsp\alpha_\ell)\alphad_\ell)}\lsp
X_3{\vphantom{X}}^{\Delta-\ell-\Delta_\phi-\Delta_R-\frac32}\,,}[tIThetaI]
and a superconformal block computation gives
\eqn{\hat{\CG}_{\Delta,\lsp\ell,\lsp\Delta_\phi-\Delta_R}^{\phib R;\,
\phi R} = \hat{c}_1\lsp g_{\Delta+\frac12,\lsp\ell}^{\Delta_\phi-\Delta_R}
+\hat{c}_2\lsp g_{\Delta+\frac32,\lsp\ell+1}^{\Delta_\phi-\Delta_R}\,,
}[blockThetaI]
where
\eqn{\begin{gathered}
\hat{c}_1=\frac{\ell+2}{(\ell+1)\big(2(\Delta-\ell-\Delta_\phi)-3\big)}\,,
\\
\hat{c}_2=\frac{(2\Delta-3)
\big(2(\Delta+\ell-\Delta_\phi)+5\big)
\big(2(\Delta+\ell+\Delta_\phi-\Delta_R)+1\big)^2}
{4(2\Delta-1)
\big(2(\Delta+\ell)+1\big)
\big(2(\Delta+\ell)+3\big)
\big(2(\Delta-\ell-\Delta_\phi)-3\big)
\big(2(\Delta+\ell+\Delta_\phi)-3\big)}\,.
\end{gathered}}[chat]
The block $\hat{\CG}_{\Delta,\lsp\ell,\lsp \Delta_\phi-\Delta_R} ^{\phib
R\lsp;\lsp \phi R}$ is another contribution to \FourPointOnes. We should
note here that if the shortening condition
$Q_{(\beta}\COb_{\alpha\alpha_1\hspace{-0.8pt}\ldots \alpha_\ell); \lsp
\alphad_1 \hspace{-0.8pt} \ldots \alphad_\ell}=0$ is satisfied, then $\CO$
is forced to have $\qb=-\frac12(\ell+1)$~\cite{Osborn:1998qu}. As a result,
the dimension of such $\CO$ is fixed to be
$\Delta=\Delta_\phi-\ell-\frac52$. This is below the unitarity bound
$\Delta\geq\Delta_\phi+\ell +\frac32$ for this class of operators, but it
nevertheless provides a check on $\hat{c}_2$ of \chat.\foot{For a general
scalar operator $\CS$ we get a block
$\hat{\CG}_{\Delta,\lsp\ell,\lsp\Delta_\phi-\Delta_S}^{\phib S;\lsp\phi
\Sb}$ similar to \blockThetaI but with
\eqn{\begin{gathered}
\hat{c}_1=\frac{\ell+2}{(\ell+1)\big(2(\Delta-\ell-\Delta_\phi
+q_\CS-\qb_\CS)-3\big)}\,,\\
\hat{c}_2=\frac{(2\Delta-3)
\big(2(\Delta+\ell-\Delta_\phi+q_\CS-\qb_\CS)+5\big)
\big(2(\Delta+\ell+\Delta_\phi-q_\CS-\qb_\CS)+1\big)^2}
{4(2\Delta-1)
\big(2(\Delta+\ell)+1\big)
\big(2(\Delta+\ell)+3\big)
\big(2(\Delta-\ell-\Delta_\phi+q_\CS-\qb_\CS)-3\big)
\big(2(\Delta+\ell+\Delta_\phi-q_\CS+\qb_\CS)-3\big)}\,.
\end{gathered}}[hatcS]
}

There is another case to consider with a $\Theta_3$, i.e.\ when we have a
superconformal primary of the form $\CO_{\alpha_1\ldots\alpha_\ell;\lsp
\alphad_2\ldots\alphad_\ell}$ for some $\ell\geq1$, again with
$q=\frac12(\Delta+\Delta_\phi-\frac32)$ and
$\qb=\frac12(\Delta-\Delta_\phi+\frac32)$. Unitarity requires
$\Delta\geq|\Delta_\phi-2|+\ell+\frac32$. Then, the conformal primary
$Q_{(\alpha_1}\COb_{\alpha_2\ldots\alpha_\ell);\lsp
\alphad_1\hspace{-0.8pt}\ldots\alphad_\ell}$ has spin $\ell$ and can
contribute to the $\phib\times R$ OPE.  Corresponding to \ThreePF we here
have
\eqn{t_{\alpha_1\hspace{-0.8pt}\ldots\alpha_\ell;\lsp\alphad_2
\hspace{-0.8pt}\ldots\alphad_\ell}(X_3)=\check{c}_{\phib R \CO_\ell}\,
\Theta_{3\lsp(\alpha_1}\text{X}_{3\lsp\alpha_2(\alphad_2}\!
\cdots\text{X}_{3\lsp\alpha_\ell)\alphad_\ell)}\lsp
X_3{\vphantom{X}}^{\Delta-\ell-\Delta_\phi-\Delta_R+\frac12}\,, \qquad
\ell\geq1\,,}[tIThetaII]
and the associated superconformal block is
\eqn{\check{\CG}_{\Delta,\lsp\ell,\lsp\Delta_\phi-\Delta_R}^ {\phib
R;\,\phi R} = \check{c}_1\lsp g_{\Delta+\frac12,
\lsp\ell}^{\Delta_\phi-\Delta_R} +\check{c}_2\lsp
g_{\Delta+\frac32,\lsp\ell-1}^{\Delta_\phi-\Delta_R}\,, \qquad
\ell\geq1\,,}[blockThetaII]
with
\eqn{\begin{gathered}
  \check{c}_1=\frac{1}{2(\Delta+\ell-\Delta_\phi)+1}\,,\\
  \check{c}_2=\frac{(\ell+1)(2\Delta-3)
  \big(2(\Delta-\ell-\Delta_\phi)+1\big)
\big(2(\Delta-\ell+\Delta_\phi-\Delta_R)-3\big)^2} {4\ell(2\Delta-1)
  \big(2(\Delta-\ell)-1\big) \big(2(\Delta-\ell)-3\big)
\big(2(\Delta+\ell-\Delta_\phi)+1\big)
\big(2(\Delta-\ell+\Delta_\phi)-7\big)}\,.  \end{gathered}}[ccheck]
For operators $\CO$ of this class such that
$Q^{\alpha}\COb_{\alpha\alpha_3\ldots\alpha_\ell;\lsp
\alphad_1\hspace{-0.8pt}\ldots\alphad_\ell}=0$, it follows that $\CO$ has
$\qb=\frac12(\ell+1)$~\cite{Osborn:1998qu}. This implies that the dimension
of such $\CO$ is $\Delta=\Delta_\phi+\ell-\frac12$, providing a check on
$\check{c}_2$ of \ccheck.\foot{For a general scalar operator $\CS$ we get a
block $\check{\CG}_{\Delta,\lsp\ell,\lsp\Delta_\phi-\Delta_S}^{\phib
S;\lsp\phi \Sb}$ similar to \blockThetaII but with
\eqn{\begin{gathered}
\check{c}_1=\frac{1}{2(\Delta+\ell-\Delta_\phi+q_\CS-\qb_\CS)+1}\,,\\
\check{c}_2=\frac{(\ell+1)(2\Delta-3)
\big(2(\Delta-\ell-\Delta_\phi+q_\CS-\qb_\CS)+1\big)
\big(2(\Delta-\ell+\Delta_\phi-q_\CS-\qb_\CS)-3\big)^2}
{4\ell(2\Delta-1)
\big(2(\Delta-\ell)-1\big) \big(2(\Delta-\ell)-3\big)
\big(2(\Delta+\ell-\Delta_\phi+q_\CS-\qb_\CS)+1\big)
\big(2(\Delta-\ell+\Delta_\phi-q_\CS+\qb_\CS)-7\big)}\,.
\end{gathered}}[checkcS]
}
Note that this dimension of $\CO$ is consistent with the unitarity
bound for this class of operators only if $\Delta_\phi\geq2$.

If $\Theta_3^2$ appears in $t_I$ only the superconformal descendant
$Q^2\COind{\alpha}$ of a superconformal primary $\COind{\alpha}$ with
$q=\frac12(\Delta+\Delta_\phi-3)$ and $\qb=\frac12(\Delta-\Delta_\phi+3)$
needs to be considered. The associated conformal block we have to include
is $g_{\Delta+1, \lsp\ell}^{\Delta_\phi-\Delta_R}$. The unitarity bound
here is $\Delta\geq|\Delta_\phi-3|+\ell+2$.

To summarize we may write, in \FourPointOnes,
\eqna{\sum_{\COb_\ell\in\phib\times R}\hspace{-4pt}|c_{\phib R\CO_\ell}|^2
\lsp\CG_{\Delta,\lsp\ell,\lsp \Delta_\phi-\Delta_R}^{\phib R\lsp;\lsp
\phi R}(u,v)&=\hspace{-4pt}
\sum_{\COb_\ell\in\phib\times R}\hspace{-4pt}|\bar{c}_{\phib R\CO_\ell}|^2
\lsp\bar{\CG}_{\Delta,\lsp\ell,\lsp \Delta_\phi-\Delta_R}^{\phib R\lsp;\lsp
\phi R}(u,v)\\
&\quad+\hspace{-8pt}\sum_{(Q\COb)_\ell\in\phib\times R}\hspace{-4pt}
|\hat{c}_{\phib R(\Qb\CO)_\ell}|^2\lsp\hat{\CG}_{\Delta,\lsp\ell,\lsp
\Delta_\phi-\Delta_R}^{\phib R\lsp;\lsp\phi R}(u,v)\\
&\quad+\hspace{-8pt}\sum_{(Q\COb)_\ell\in\phib\times R}\hspace{-4pt}
|\check{c}_{\phib R(\Qb\CO)_\ell}|^2\lsp\check{\CG}_{\Delta,\lsp\ell,\lsp
\Delta_\phi-\Delta_R}^{\phib R\lsp;\lsp\phi R}(u,v)\\
&\quad+\hspace{-8pt}\sum_{(Q^2\COb)_\ell\in\phib\times R}\hspace{-4pt}
|c_{\phib R (\Qb^2\CO)_\ell}|^2\lsp
g_{\Delta+1,\lsp\ell}^{\Delta_\phi-\Delta_R}(u,v)\,,}[]
with the appropriate unitarity bounds, and with the contribution associated
to \ThreePFPhi implicitly included in the first sum on the right-hand side.

Let us finally consider $\vev{\phib(x_1)\lsp R(x_2)\lsp R(x_3)\lsp
\phi(x_4)}$ both in the $12\rightarrow34$ and the $14\rightarrow32$
channel.  For the former we have
\eqna{\vev{\phib(x_1)\lsp R(x_2)\lsp R(x_3)\lsp \phi(x_4)}
&=\smash{\frac{1}{r_{12}^{\,\Delta_\phi+\Delta_R}\,r_{34}
^{\,\Delta_\phi+\Delta_R}}
\left(\frac{r_{13}\lsp r_{24}}{r_{14}^{\lsp 2}}\right)^{\Delta_\phi
-\Delta_R}}\\
&\hspace{5cm}\smash{\times\hspace{-4pt}\sum_{\COb_\ell\in\phib\times R}
\hspace{-4pt}|c_{\phib R\CO_\ell}|^2\lsp(-1)^\ell\lsp
\CG_{\Delta,\lsp\ell,\lsp\Delta_\phi-\Delta_R}^{\phib R\lsp;\lsp
R\phi}(u,v)}\,,}[FourPointOnesII]
where one contribution comes from
\eqn{\bar{\CG}_{\Delta,\lsp\ell,\lsp \Delta_\phi-\Delta_R}
^{\phib R\lsp;\lsp R\phi}=
\tilde{g}_{\Delta,\lsp\ell}^{\Delta_\phi-\Delta_R}
-\bar{c}_1\,\tilde{g}_{\Delta+1,\lsp\ell+1}^{\Delta_\phi-\Delta_R}
-\bar{c}_2\,\tilde{g}_{\Delta+1,\lsp\ell-1}^{\Delta_\phi-\Delta_R}
+\bar{c}_1 \bar{c}_2\,\tilde{g}_{\Delta+2,\lsp\ell}
^{\Delta_\phi-\Delta_R}\,,\qquad
\tilde{g}_{\alpha,\lsp\beta}^{\gamma}\equiv
g_{\alpha,\lsp\beta}^{\gamma,\lsp-\gamma}\,.}[SCBOneII]
As before, there are also contributions corresponding to superconformal
descendants whose primary does not appear in the $\phib\times R$ OPE. In
particular, corresponding to \blockThetaI and \blockThetaII we have
\eqn{\hat{\CG}_{\Delta,\lsp\ell,\lsp\Delta_\phi-\Delta_R}^{\phib R;\,
R\phi} = \hat{c}_1\lsp \tilde{g}_{\Delta+\frac12,\lsp\ell}^
{\Delta_\phi-\Delta_R}
-\hat{c}_2\lsp \tilde{g}_{\Delta+\frac32,\lsp\ell+1}^{\Delta_\phi
-\Delta_R}\,,}[blockThetaXI]
and
\eqn{\check{\CG}_{\Delta,\lsp\ell,\lsp\Delta_\phi-\Delta_R}^
{\phib R;\,R\phi} = \check{c}_1\lsp \tilde{g}_{\Delta+\frac12,
\lsp\ell}^{\Delta_\phi-\Delta_R}
-\check{c}_2\lsp \tilde{g}_{\Delta+\frac32,\lsp\ell-1}^{\Delta_\phi
-\Delta_R}\,,\qquad \ell\geq1\,,}[blockThetaXII]
while we also have the $\tilde{g}_{\Delta+1,\lsp\ell}^{\Delta_\phi
-\Delta_R}$ conformal block contribution. The unitarity bounds are as
explained above.

In the $14\rightarrow32$ channel we can use results
of~\cite{Khandker:2014mpa} to obtain
\eqn{\vev{\phib(x_1)\lsp R(x_2)\lsp R(x_3)\lsp \phi(x_4)}
=\frac{1}{r_{14}^{\,2\Delta_\phi}\,r_{23}^{\,2\Delta_R}}
\sum_{\substack{\CO_\ell\in\phib\times \phi\\\CO_\ell\in
R\times R}}\hspace{-4pt}\lsp(-1)^\ell\lsp\CG_{\Delta,\lsp\ell}^{\phib
\phi;\lsp RR}(v,u)\,,}[FourPointTwos]
where
\eqn{\CG_{\Delta,\lsp\ell \text{ even}}^{\phib\phi;\lsp RR}=
c^\ast_{\phib\phi\CO_\ell}c^{\lsp (0)}_{RR\CO_\ell}\, g_{\Delta,\lsp\ell}
-\frac{c^\ast_{\phib\phi\CO_\ell}\left((\Delta+\ell)^2
c^{\lsp (0)}_{RR\CO_\ell}
-8(\Delta-1)c^{\lsp (2)}_{RR\CO_\ell}\right)}
{16\Delta(\Delta-\ell-1)(\Delta+\ell+1)}\,
g_{\Delta+2,\lsp\ell}\,,}[evBl]
and
\eqn{\CG_{\Delta,\lsp\ell \text{ odd}}^{\phib\phi;\lsp RR}=
-\frac{c^\ast_{\phib\phi\CO_\ell}c^{\lsp (1)}_{RR\CO_\ell}}
{2(\Delta+\ell+1)}\,g_{\Delta+1,\lsp\ell+1}
-\frac{c^\ast_{\phib\phi\CO_\ell}\left(c^{\lsp (1)}_{RR\CO_\ell}
+\frac{\ell+1}{\ell}c^{\lsp (3)}_{RR\CO_\ell}\right)}
{2(\Delta-\ell-1)}
\,g_{\Delta+1,\lsp\ell-1}\,.}[oddBl]

\subsec{Four-point function \texorpdfstring{$\langle R(x_1)\lsp R(x_2)\lsp
R(x_3)\lsp R(x_4)\rangle$}{<RRRR>}}
In the $12\rightarrow34$ channel we can write
\eqn{\vev{R(x_1)\lsp R(x_2)\lsp R(x_3)\lsp R(x_4)}=
\frac{1}{r_{12}^{\,2\Delta_R}\, r_{34}^{\,2\Delta_R}}
\sum_{\CO_\ell\in R\times R}\hspace{-4pt}
\CG_{\Delta,\lsp\ell}^{RR\lsp;\lsp RR}(u,v)\,.}[RRRR]
Here the sum runs over superconformal primaries, but also over just
conformal primaries if a superconformal primary does not contribute but one
of its descendants does. Only even-spin operators can be exchanged in the
$R\times R$ OPE. These can come from even- or odd-spin superconformal
primaries, so that the sum in \RRRR runs over $\CO_\ell$'s with both even
and odd spin. The block $G_{\Delta,\lsp\ell}^{RR\lsp;\lsp RR}$, then,
receives separate contributions from even- and odd-spin superconformal
primaries. There are no constraints on $R$, except that it is a real
operator of dimension $\Delta\geq\ell+2$ by unitarity, and so from results
of~\cite{Khandker:2014mpa} we see that we cannot fix the coefficients of
the conformal block contributions to the superconformal blocks. The best we
can do is write
\eqn{\CG_{\Delta,\lsp\ell \text{ even}}^{RR\lsp;\lsp RR}=
|c^{\lsp (0)}_{RR\CO_\ell}|^2 g_{\Delta,\lsp\ell}
+\frac{\left|(\Delta+\ell)^2c^{\lsp (0)}_{RR\CO_\ell}
-8(\Delta-1)c^{\lsp (2)}_{RR\CO_\ell}\right|^2}
{16\Delta^2(\Delta-\ell-1) (\Delta-\ell-2)
(\Delta+\ell)(\Delta+\ell+1)}\,
g_{\Delta+2,\lsp\ell}\,,}[RRBlockEven]
and
\eqn{\CG_{\Delta,\lsp\ell \text{ odd}}^{RR\lsp;\lsp RR}=
\frac{|c^{\lsp (1)}_{RR\CO_\ell}|^2}{(\Delta+\ell)
(\Delta+\ell+1)}\,g_{\Delta+1,\lsp\ell+1}
+\frac{\left|c^{\lsp (1)}_{RR\CO_\ell}
+\frac{\ell+1}{\ell}c^{\lsp (3)}_{RR\CO_\ell}\right|^2}
{(\Delta-\ell-1)(\Delta+\ell+1)}
\,g_{\Delta+1,\lsp\ell-1}\,.}[RRBlockOdd]

A superconformal primary that is not an integer-spin Lorentz representation
can have superconformal descendant conformal primary components that
contribute to \RRRR. It turns out that we only need to consider
superconformal primaries of the form $\CO_{\alpha\alpha_1
\hspace{-0.8pt}\ldots\alpha_\ell;\lsp \alphad_2\hspace{-0.8pt}
\ldots\alphad_\ell}$ with even $\ell\geq2$ and
$q=\qb=\frac12\Delta$.\foot{The three-point function
$\vev{\CR(z_1)\lsp\CR(z_2)\lsp\CO_I(z_3)}$ is symmetric under
$z_1\leftrightarrow z_2$, something that restricts the possible
non-integer-spin superconformal primary operators we can consider. We thank
Ran Yacoby for discussions on this point.} The relevant operator is then
the conformal primary contained in the superconformal descendant
$\Qb_{(\alphad_1}Q^\alpha\CO_{\alpha\alpha_1\hspace{-0.8pt}
\ldots\alpha_\ell;\lsp  \alphad_2\ldots\alphad_\ell)}$, where the undotted
indices are the only ones that are symmetrized with $\alphad_1$. The
conformal block we need to include is $g_{\Delta+1,\lsp\ell}$ with even
$\ell\geq2$ and $\Delta\geq \ell+3$ by unitarity.

\newsec{Four-point functions with linear multiplets}[secJBl]
So far we have analyzed four-point functions including a chiral operator
$\phi$, its conjugate $\phib$, and a real field $R$. The results we have
obtained can be easily adapted to the case where the corresponding real
superfield $\CR$ is a linear multiplet $\CJ$, containing a $\text{U}(1)$
vector current $j^\mu$. Linear multiplets have $q_\CJ=\qb_\CJ=1$, and
appear in theories with global symmetries. The superspace three-point
function $\langle\CJ(z_1)\lsp\CJ(z_2)\lsp\CO(z_3)\rangle$ was considered
in~\cite{Fortin:2011nq}, where the superconformal blocks for $\langle
J(x_1)\lsp J(x_2) \lsp J(x_3)\lsp J(x_4)\rangle$ were computed. Bootstrap
constraints from $\langle J(x_1)\lsp J(x_2)\lsp J(x_3)\lsp J(x_4) \rangle$
were obtained in~\cite{Berkooz:2014yda}. Our aim here is to obtain bounds
using the system of correlators $\langle\phib(x_1)\lsp \phi(x_2)\lsp
\phib(x_3)\lsp \phi(x_4)\rangle$, $\langle \phib(x_1)\lsp J(x_2)\lsp
\phi(x_3)\lsp J(x_4)\rangle$, and $\langle J(x_1)\lsp J(x_2)\lsp J(x_3)\lsp
J(x_4)\rangle$.

The associated superconformal-block decomposition of these four-point
functions can be obtained from the results of section \secSCB, given that
$\CJ$ is a particular case of a real superfield with $q_\CJ=\qb_\CJ=1$.
Since $Q^2(J)=\Qb^2(J)=0$ and $Q_\alpha(\phib)=0$, we also need to make
sure that the operators in the right hand side of the $\phib\times J$ OPE
are annihilated by $Q^2$.  This last requirement implies that a
superconformal primary of the form $\COind{\alpha}$, as considered around
\tI above, can only have $\qb=1$ and $\ell=0$~\cite{Osborn:1998qu}, i.e.\
it can be a scalar with $\Delta=\Delta_\phi+2$. This implies that,
analogously to the blocks defined in \SCBOne and \SCBOneII, we only need
\eqn{\bar{\CG}_{\Delta_\phi+2,\lsp 0,\lsp \Delta_\phi}^{\phib J;\lsp\phi
J}=g_{\Delta_\phi+2,\lsp 0}^{\Delta_\phi-2}\,,\qquad
\bar{\CG}_{\Delta_\phi+2,\lsp0,\lsp \Delta_\phi}^{\phib J;\lsp
J\phi}=\tilde{g}_{\Delta_\phi+2,\lsp 0}^{\Delta_\phi-2}\,.}[]
Without any changes other than $\Delta_R\to\Delta_J=2$ we can define
$\hat{\CG}_{\Delta,\lsp\ell,\lsp \Delta_\phi}^{\phib J;\lsp\phi J}$,
$\hat{\CG}_{\Delta,\lsp\ell,\lsp \Delta_\phi}^{\phib J;\lsp J\phi}$,
$\check{\CG}_{\Delta,\lsp\ell,\lsp \Delta_\phi}^{\phib J;\lsp\phi J}$, and
$\check{\CG}_{\Delta,\lsp\ell,\lsp \Delta_\phi}^{\phib J;\lsp J\phi}$ using
\blockThetaI, \blockThetaXI, \blockThetaII, and \blockThetaXII,
respectively, as well as $g_{\Delta+1, \lsp\ell}^{\Delta_\phi-2}$ with
$\Delta\geq|\Delta_\phi-3|+\ell+2$.

For the blocks defined in \evBl, \oddBl, \RRBlockEven, and \RRBlockOdd we
need to use relations that exist between $c_{JJ\CO_\ell}^{(2)}$ and
$c_{JJ\CO_\ell}^{(0)}$, as well as between $c_{JJ\CO\ell}^{(3)}$ and
$c_{JJ\CO_\ell}^{(1)}$, namely~\cite{Khandker:2014mpa}
\eqn{c_{JJ\CO_\ell}^{(2)}=-\tfrac18(\Delta+\ell)(\Delta-\ell-4)
c_{JJ\CO_\ell}^{(0)}\,,\qquad
c_{JJ\CO_\ell}^{(3)}=-\frac{2(\Delta-2)}{\Delta+\ell}
c_{JJ\CO_\ell}^{(1)}\,.}[]
Using this we can define, in the $14\rightarrow32$ channel,
\eqn{\vev{\phib(x_1)\lsp J(x_2)\lsp J(x_3)\lsp \phi(x_4)}
=\frac{1}{r_{14}^{\,2\Delta_\phi}\,r_{23}^{\,4}}
\sum_{\substack{\CO_\ell\in\phib\times \phi\\\CO_\ell\in
J\times J}}\hspace{-4pt}c_{\phib\phi\CO_\ell}^\ast c_{JJ\CO_\ell}\lsp
(-1)^\ell\lsp\CG_{\Delta,\lsp\ell}^{\phib \phi;\lsp JJ}(v,u)\,,
}[FourPointJTwos]
where
\eqn{\CG_{\Delta,\lsp\ell \text{ even}}^{\phib\phi;\lsp JJ}=
g_{\Delta,\lsp\ell}
-\frac{(\Delta-2)(\Delta+\ell)(\Delta-\ell-2)}
{16\Delta(\Delta-\ell-1)(\Delta+\ell+1)}\,
g_{\Delta+2,\lsp\ell}\,,}[evJBl]
and
\eqn{\CG_{\Delta,\lsp\ell \text{ odd}}^{\phib\phi;\lsp JJ}=
-\frac{1}{2(\Delta+\ell+1)}\,g_{\Delta+1,\lsp\ell+1}
+\frac{(\ell+2)(\Delta-\ell-2)}{2\ell(\Delta+\ell)(\Delta-\ell-1)}
\,g_{\Delta+1,\lsp\ell-1}\,.}[oddJBl]
Finally, in the $12\rightarrow34$ channel we can write
\eqn{\vev{J(x_1)\lsp J(x_2)\lsp J(x_3)\lsp J(x_4)}=
\frac{1}{r_{12}^{\,4}\, r_{34}^{\,4}}
\sum_{\CO_\ell\in J\times J}\hspace{-4pt}|c_{RR\CO_\ell}|^2\lsp
\CG_{\Delta,\lsp\ell}^{JJ;\lsp JJ}(u,v)\,,}[JJJJ]
with
\eqn{\CG_{\Delta,\lsp\ell \text{ even}}^{JJ;\lsp JJ}=
g_{\Delta,\lsp\ell}
+\frac{(\Delta-2)^2(\Delta+\ell)(\Delta-\ell-2)}
{16\Delta^2(\Delta-\ell-1)(\Delta+\ell+1)}\,
g_{\Delta+2,\lsp\ell}\,,}[JJBlockEven]
and
\eqn{\CG_{\Delta,\lsp\ell \text{ odd}}^{JJ;\lsp JJ}=
\frac{1}{(\Delta+\ell)(\Delta+\ell+1)}\,g_{\Delta+1,\lsp\ell+1}
+\frac{(\ell+2)^2(\Delta-\ell-2)}
{\ell^2(\Delta+\ell)^2(\Delta-\ell-1)}
\,g_{\Delta+1,\lsp\ell-1}\,.}[JJBlockOdd]

We should also mention here that there are conformal primary superconformal
descendant operators that contribute to the four-point functions involving
$J$, but whose corresponding superconformal primaries do not. This type of
operators has been analyzed in detail in~\cite{Berkooz:2014yda}. The result
is that in order to account for these operators we need to include
$g_{\Delta+1,\lsp\ell}$ with even $\ell\geq2$ and $\Delta\geq \ell+3$ by
unitarity.

\newsec{Crossing relations}[crossRel]
Using the results of section \secSCB  we can now write down the crossing
equations that we use in our numerical analysis. It is well-known that from
$\vev{\phib(x_1)\lsp \phi(x_2)\lsp \phib(x_3)\lsp \phi(x_4)}$ we obtain
three crossing relations~\cite{Poland:2011ey}. We get another three from
$\vev{\phib(x_1)\lsp R(x_2)\lsp \phi(x_3)\lsp R(x_4)}$ (for these we will
assume that $1\leq\Delta_\phi<2$), and a final crossing relation from
$\vev{R(x_1)\lsp R(x_2)\lsp R(x_3)\lsp R(x_4)}$.  In total we have seven
crossing relations.

\subsection{Chiral-chiral and chiral-antichiral}
From $\vev{\phib(x_1)\lsp \phi(x_2)\lsp \phib(x_3)\lsp \phi(x_4)}$
we find the crossing relations~\cite{Poland:2011ey}
\eqn{\sum_{\CO_\ell\in\phib\times\phi}
|c_{\phib\phi\CO_\ell}|^2
\begin{pmatrix}
  \CF^{\lsp\phib\phi;\lsp \phi\phib}_{\Delta,\lsp\ell,\lsp\Delta_\phi}(u,v)
  \vspace{5pt}\\
  \CH^{\phib\phi;\lsp \phi\phib}_{\Delta,\lsp\ell,\lsp \Delta_\phi}
  (u,v)\vspace{5pt}\\
  (-1)^\ell\CF^{\lsp\phib\phi;\lsp \phib\phi}_{\Delta,\lsp\ell,
  \lsp \Delta_\phi}(u,v)
\end{pmatrix}\hspace{5pt}
+\hspace{-4pt}\sum_{\COb_\ell\in\phib\times\phib}
|c_{\phib\phib\CO_\ell}|^2
\begin{pmatrix}
  F^{\lsp\phib\phib;\lsp \phi\phi}_{\Delta,\lsp\ell,\lsp \Delta_\phi}(u,v)
  \vspace{5pt}\\
  -H^{\phib\phib;\lsp \phi\phi}_{\Delta,\lsp\ell,\lsp \Delta_\phi}(u,v)
  \vspace{5pt}\\
  0
\end{pmatrix}=0\,,}[crossRelphi]
where
\eqna{\CF^{\lsp\phib\phi;\lsp\phi\phib}_{\Delta,\lsp\ell,\lsp\Delta_\phi}
(u,v)&=u^{-\Delta_\phi}\CG^{\lsp\phib\phi;\lsp\phi\phib}_{\Delta,\lsp\ell}
(u,v)-(u\leftrightarrow v)\,,\\
\CH^{\phib\phi;\lsp\phi\phib}_{\Delta,\lsp\ell,\lsp\Delta_\phi}
(u,v)&=u^{-\Delta_\phi}\CG^{\lsp\phib\phi;\lsp\phi\phib}_{\Delta,\lsp\ell}
(u,v)+(u\leftrightarrow v)\,,\\
\CF^{\lsp\phib\phi;\lsp\phib\phi}_{\Delta,\lsp\ell,\lsp\Delta_\phi}
(u,v)&=u^{-\Delta_\phi}\CG^{\lsp\phib\phi;\lsp\phib\phi}_{\Delta,\lsp\ell}
(u,v)-(u\leftrightarrow v)\,,\\
F^{\lsp\phib\phib;\lsp\phi\phi}_{\Delta,\lsp\ell,\lsp\Delta_\phi}
(u,v)&=u^{-\Delta_\phi}g_{\Delta,\lsp\ell}
(u,v)-(u\leftrightarrow v)\,,\\
H^{\phib\phib;\lsp\phi\phi}_{\Delta,\lsp\ell,\lsp\Delta_\phi}
(u,v)&=u^{-\Delta_\phi}g_{\Delta,\lsp\ell}
(u,v)+(u\leftrightarrow v)\,.}[Fphi]

\subsection{Chiral-real}
From $\vev{\phib(x_1)\lsp R(x_2)\lsp R(x_3)\lsp \phi(x_4)}$ we find
\eqna{\sum_{\COb_\ell\in\phib\times R}\hspace{-4pt}
|\bar{c}_{\phib R\CO_\ell}|^2\lsp(-1)^\ell\lsp\bar{\CF}_{\Delta,\lsp\ell,
\lsp\Delta_\phi,\lsp\Delta_R}^{\lsp\phib R\lsp;\lsp R\phi}
&+\hspace{-8pt}\sum_{(Q\COb)_\ell\in\phib\times R}
\hspace{-6pt}|\hat{c}_{\phib R(\Qb\CO)_\ell}|^2\lsp(-1)^\ell\lsp
\hat{\CF}^{\lsp\phib R\lsp;\lsp R\phi}_{\Delta,\lsp\ell,
\lsp\Delta_\phi,\lsp\Delta_R}\\
&\hspace{-4.5cm}+\hspace{-8pt}\sum_{(Q\COb)_\ell\in\phib\times R}
\hspace{-6pt}|\check{c}_{\phib R(\Qb\CO)_\ell}|^2\lsp(-1)^\ell\lsp
\check{\CF}^{\lsp\phib R\lsp;\lsp R\phi}_{\Delta,\lsp\ell,
\lsp\Delta_\phi,\lsp\Delta_R}
+\hspace{-8pt}\sum_{(Q^2\COb)_\ell\in\phib\times R}
\hspace{-6pt}|c_{\phib R(\Qb^2\CO)_\ell}|^2\lsp(-1)^\ell\lsp
F^{\lsp\phib R\lsp;\lsp R\phi}_{\Delta,\lsp\ell,
\lsp\Delta_\phi,\lsp\Delta_R}\\
&\hspace{2.5cm}+\hspace{-4pt}\sum_{\CO_\ell\in\phib\times\phi}\hspace{-4pt}
c^\ast_{\phib\phi\CO_\ell}
c_{RR\CO_\ell}\lsp(-1)^\ell\lsp\CF_{\Delta,\lsp\ell,\lsp\Delta_R}
^{\lsp\phib\phi;\lsp RR}=0\,,}[crossRelphiRI]
and
\eqna{\sum_{\COb_\ell\in\phib\times R}\hspace{-4pt}
|\bar{c}_{\phib R\CO_\ell}|^2\lsp(-1)^\ell\lsp\bar{\CH}_{\Delta,\lsp\ell,
\lsp\Delta_\phi,\lsp\Delta_R}^
{\phib R\lsp;\lsp R\phi}
&+\hspace{-8pt}\sum_{(Q\COb)_\ell\in\phib\times R}
\hspace{-6pt}|\hat{c}_{\phib R(\Qb\CO)_\ell}|^2\lsp(-1)^\ell\lsp
\hat{\CH}^{\phib R\lsp;\lsp R\phi}_{\Delta,\lsp\ell,
\lsp\Delta_\phi,\lsp\Delta_R}\\
&\hspace{-4.5cm}+\hspace{-8pt}\sum_{(Q\COb)_\ell\in\phib\times R}
\hspace{-6pt}|\check{c}_{\phib R(\Qb\CO)_\ell}|^2\lsp(-1)^\ell\lsp
\check{\CH}^{\phib R\lsp;\lsp R\phi}_{\Delta,\lsp\ell,
\lsp\Delta_\phi,\lsp\Delta_R}
+\hspace{-8pt}\sum_{(Q^2\COb)_\ell\in\phib\times R}
\hspace{-6pt}|c_{\phib R(\Qb^2\CO)_\ell}|^2\lsp(-1)^\ell\lsp
H^{\phib R\lsp;\lsp R\phi}_{\Delta,\lsp\ell,
\lsp\Delta_\phi,\lsp\Delta_R}\\
&\hspace{2.5cm}-\hspace{-4pt}\sum_{\CO_\ell\in\phib\times\phi}\hspace{-4pt}
c^\ast_{\phib\phi\CO_\ell}
c_{RR\CO_\ell}\lsp(-1)^\ell\lsp\CH_{\Delta,\lsp\ell,\lsp\Delta_R}
^{\phib\phi;\lsp RR}=0\,,}[crossRelphiRII]
where
\eqna{\bar{\CF}_{\Delta,\lsp\ell,\lsp
\Delta_\phi,\lsp\Delta_R}^{\lsp\phib R\lsp;\lsp R\phi}(u,v)&=
u^{-\frac12(\Delta_\phi+\Delta_R)}\lsp\bar{\CG}_{\lsp\Delta,\lsp\ell,\lsp
\Delta_\phi-\Delta_R}^{\phib R\lsp;\lsp R \phi}(u,v) -
(u\leftrightarrow v)\,,\\
\bar{\CH}_{\Delta,\lsp\ell,\lsp
\Delta_\phi,\lsp\Delta_R}^{\phib R\lsp;\lsp R\phi}(u,v)&=
u^{-\frac12(\Delta_\phi+\Delta_R)}\lsp\bar{\CG}_{\lsp\Delta,\lsp\ell,\lsp
\Delta_\phi-\Delta_R}^{\phib R\lsp;\lsp R \phi}(u,v) +
(u\leftrightarrow v)\,,}[FphiRII]
and similarly for $\hat{\CF},\hat{\CH},\check{\CF},\check{\CH}$, using
$\hat{\CG},\check{\CG}$,
\eqna{F_{\Delta,\lsp\ell,\lsp
\Delta_\phi,\lsp\Delta_R}^{\lsp\phib R\lsp;\lsp R\phi}(u,v)&=
u^{-\frac12(\Delta_\phi+\Delta_R)}\lsp \tilde{g}_{\lsp\Delta,\lsp\ell,
\lsp\Delta_\phi-\Delta_R}^{\phib R\lsp;\lsp R \phi}(u,v) -
(u\leftrightarrow v)\,,\\
H_{\Delta,\lsp\ell,\lsp
\Delta_\phi,\lsp\Delta_R}^{\phib R\lsp;\lsp R\phi}(u,v)&=
u^{-\frac12(\Delta_\phi+\Delta_R)}\lsp\tilde{g}_{\lsp\Delta,\lsp\ell,
\lsp\Delta_\phi-\Delta_R}^{\phib R\lsp;\lsp R \phi}(u,v) +
(u\leftrightarrow v)\,,}[]
and, if $\ell$ is even, $c_{RR\CO_\ell}=c_{RR\CO_\ell}^{(0)}$ and
\eqna{\CF_{\Delta,\lsp\ell,\lsp\Delta_R}^{\lsp\phib\phi;\lsp RR}(u,v)&=
u^{-\Delta_R}\lsp\CG_{\lsp\Delta,\lsp\ell\text{ even}}^{\phib\phi;
\lsp RR}(u,v) - (u\leftrightarrow v)\,,\\
\CH_{\Delta,\lsp\ell,\lsp\Delta_R}^{\phib\phi;\lsp RR}(u,v)&=
u^{-\Delta_R}\lsp\CG_{\lsp\Delta,\lsp\ell\text{ even}}^{\phib\phi;
\lsp RR}(u,v) +(u\leftrightarrow v)\,,}[rescBlI]
while, if $\ell$ is odd, $c_{RR\CO_\ell}=c_{RR\CO_\ell}^{(1)}$ and
\eqna{\CF_{\Delta,\lsp\ell,\lsp\Delta_R}^{\lsp\phib\phi;\lsp RR}(u,v)&=
u^{-\Delta_R}\lsp\CG_{\lsp\Delta,\lsp\ell\text{ odd}}^{\phib\phi;
\lsp RR}(u,v) - (u\leftrightarrow v)\,,\\
\CH_{\Delta,\lsp\ell,\lsp\Delta_R}^{\phib\phi;\lsp RR}(u,v)&=
u^{-\Delta_R}\lsp\CG_{\lsp\Delta,\lsp\ell\text{ odd}}^{\phib\phi;
\lsp RR}(u,v) +(u\leftrightarrow v)\,.}[rescBlII]
Note that in \rescBlI and \rescBlII the superconformal blocks of \evBl and
\oddBl have been rescaled by
$c^\ast_{\phib\phi\CO_\ell}c_{RR\CO_{\ell}}^{(0)}$ and
$c^\ast_{\phib\phi\CO_\ell}c_{RR\CO_{\ell}}^{(1)}$, respectively.

The crossing relation arising from $\vev{\phib(x_1)\lsp R(x_2)\lsp
\phi(x_3)\lsp R(x_4)}$ is
\eqna{\sum_{\COb_\ell\in\phib\times R}
\hspace{-4pt}|\bar{c}_{\phib R\CO_\ell}|^2
\lsp\bar{\CF}_{\lsp\Delta,\lsp\ell,\lsp
\Delta_\phi,\lsp\Delta_R}^{\lsp\phib R\lsp;\lsp \phi R}
&+\hspace{-8pt}\sum_{(Q\COb)_\ell\in\phib\times R}
\hspace{-6pt}|\hat{c}_{\phib R(\Qb\CO)_\ell}|^2
\hat{\CF}^{\lsp\phib R\lsp;\lsp \phi R}_{\Delta,\lsp\ell,\lsp\Delta_\phi,
\lsp\Delta_R}\\
&\hspace{-1.5cm}+\hspace{-8pt}\sum_{(Q\COb)_\ell\in\phib\times R}
\hspace{-6pt}|\check{c}_{\phib R(\Qb\CO)_\ell}|^2
\check{\CF}^{\lsp\phib R\lsp;\lsp \phi R}_{\Delta,\lsp\ell,
\lsp\Delta_\phi,\lsp\Delta_R}
+\hspace{-8pt}\sum_{(Q^2\COb)_\ell\in\phib\times R}
\hspace{-6pt}|c_{\phib R(\Qb^2\CO)_\ell}|^2
F^{\lsp\phib R\lsp;\lsp \phi R}_{\Delta,\lsp\ell,\lsp\Delta_\phi,
\lsp\Delta_R}
=0\,,}[crossRelphiRIII]
where
\eqna{\bar{\CF}_{\Delta,\lsp\ell,\lsp
\Delta_\phi,\lsp\Delta_R}^{\lsp\phib R\lsp;\lsp \phi R}(u,v)&=
u^{-\frac12(\Delta_\phi+\Delta_R)}\lsp\bar{\CG}_{\lsp\Delta,\lsp\ell,\lsp
\Delta_\phi-\Delta_R}^{\phib R\lsp;\lsp \phi R}(u,v) - (u\leftrightarrow
v)\,,\\
F_{\Delta,\lsp\ell,\lsp
\Delta_\phi,\lsp\Delta_R}^{\lsp\phib R\lsp;\lsp \phi R}(u,v)&=
u^{-\frac12(\Delta_\phi+\Delta_R)}\lsp g^{\Delta_\phi-\Delta_R}
_{\lsp\Delta,\lsp\ell}(u,v) - (u\leftrightarrow v)\,,}[FphiR]
and similarly for $\hat{\CF},\check{\CF}$.

\subsection{Real-real}
From $\vev{R(x_1)\lsp R(x_2)\lsp R(x_3)\lsp R(x_4)}$ we find the crossing
relation
\eqn{\sum_{\CO_\ell\in R\times R}|c_{RR\CO_\ell}|^2\lsp
\CF_{\Delta,\lsp\ell,\lsp\Delta_R}^{\lsp RR\lsp;\lsp RR}\,\,
+\hspace{-10pt}\sum_{(Q\CO)_\ell\in R\times R}\hspace{-6pt}
|c_{RR(Q\CO)_\ell}|^2 F_{\Delta,\lsp\ell,\lsp\Delta_R}^{\lsp RR;\lsp
RR}=0\,, }[crossRelRR]
with
\eqna{\CF_{\Delta,\lsp\ell,\lsp\Delta_R}^{\lsp RR\lsp;\lsp RR}(u,v)&=
u^{-\Delta_R}\CG_{\Delta,\lsp\ell}^{RR\lsp;\lsp
RR}(u,v)-(u\leftrightarrow v)\,,\\
F_{\Delta,\lsp\ell,\lsp\Delta_R}^{\lsp RR\lsp;\lsp RR}(u,v)&=
u^{-\Delta_R} g_{\Delta,\lsp\ell}(u,v)-(u\leftrightarrow v)\,,}[FRR]
and for $\ell$ even we define $c_{RR\CO_\ell}=c_{RR\CO_\ell}^{(0)}$ and use
\RRBlockEven rescaled by $|c_{RR\CO_\ell}^{(0)}|^2$, while for $\ell$ odd
we define $c_{RR\CO_\ell}=c_{RR\CO_\ell}^{(1)}$ and use \RRBlockOdd
rescaled by $|c_{RR\CO_\ell}^{(1)}|^2$.

\subsection{System of crossing relations}
The crossing relations \crossRelphi, \crossRelphiRI, \crossRelphiRII,
\crossRelphiRIII and \crossRelRR can now be written in the form
\eqna{&\sum_{\substack{\CO_\ell\in\phib\times\phi\\\CO_\ell\in R\times R}}
\hspace{-4pt}\begin{pmatrix}
  c^\ast_{\phib\phi\CO_\ell} & c^\ast_{RR\CO_\ell} &
  c^{\lsp\prime\lsp\ast}_{RR\CO_\ell}
\end{pmatrix}
\vec{V}_{\Delta,\lsp\ell,\lsp\Delta_\phi,\lsp\Delta_R}
\begin{pmatrix}
  c_{\phib\phi\CO_\ell}\\ c_{RR\CO_\ell}\\ c^{\lsp\prime}_{RR\CO_\ell}
\end{pmatrix}\,\,
+\hspace{-4pt}\sum_{\COb_\ell\in\phib\times\phib}\hspace{-4pt}|c_{\phib\phib\CO_\ell}|^2
\vec{W}_{\Delta,\lsp\ell,\lsp\Delta_\phi}\\
&\quad+\hspace{-4pt}\sum_{\COb_\ell\in\phib\times R}\hspace{-4pt}
|\bar{c}_{\phib R\CO_\ell}|^2\vec{\bar{X}}_{\Delta,\lsp\ell,\lsp
\Delta_\phi,\lsp\Delta_R}
+\hspace{-8pt}\sum_{(Q\COb)_\ell\in\phib\times R}\hspace{-8pt}
|\hat{c}_{\phib R(\Qb\CO)_\ell}|^2\lsp\vec{\hat{X}}_{\Delta,\lsp\ell,
\lsp\Delta_\phi,\lsp\Delta_R}
+\hspace{-8pt}\sum_{(Q\COb)_\ell\in\phib\times R}
\hspace{-8pt}|\check{c}_{\phib R(\Qb\CO)_\ell}|^2\lsp
\vec{\check{X}}_{\Delta,\lsp\ell,\lsp\Delta_\phi,\lsp\Delta_R}
\\
&\hspace{4.8cm}+\hspace{-8pt}\sum_{(Q^2\COb)_\ell\in\phib\times R}
\hspace{-8pt}|c_{\phib R(\Qb^2\CO)_\ell}|^2\lsp
\vec{Y}_{\Delta,\lsp\ell,\lsp\Delta_\phi,\lsp\Delta_R}
+\hspace{-8pt}\sum_{(Q\CO)_\ell\in R\times R}\hspace{-8pt
}|c_{RR(Q\CO)_\ell}|^2\vec{Z}_{\Delta,\lsp\ell,\lsp\Delta_R}=0\,,
}[CrossVec]
where the seven-vector
$\vec{V}_{\Delta,\lsp\ell,\lsp\Delta_\phi,\lsp\Delta_R}$ contains the
$3\times 3$ matrices
\begin{gather*}
V^1_{\Delta,\lsp\ell,\lsp\Delta_\phi}=\begin{pmatrix}
\CF^{\lsp\phib\phi;\lsp
\phi\phib}_{\Delta,\lsp\ell,\lsp\Delta_\phi} & 0 & 0\\
0 & 0 & 0\\
0 & 0 & 0
\end{pmatrix}\,,\qquad
V^2_{\Delta,\lsp\ell,\lsp\Delta_\phi}=\begin{pmatrix}
\CH^{\phib\phi;\lsp
\phi\phib}_{\Delta,\lsp\ell,\lsp\Delta_\phi} & 0 & 0\\
0 & 0 & 0\\
0 & 0 & 0
\end{pmatrix}\,,\\
V^3_{\Delta,\lsp\ell,\lsp\Delta_\phi}=\begin{pmatrix}
(-1)^\ell\CF^{\lsp\phib\phi;\lsp
\phib\phi}_{\Delta,\lsp\ell,\lsp\Delta_\phi} & 0 & 0\\
0 & 0 & 0\\
0 & 0 & 0
\end{pmatrix}\,,\\\displaybreak[0]
V^4_{\Delta,\lsp\ell,\lsp\Delta_R}=\begin{pmatrix}
  0 & \tfrac12\lsp(-1)^{\ell}\lsp F_{1,\lsp\Delta,\lsp\ell,\lsp\Delta_R}^{\lsp\phib\phi;\lsp RR}
  & \tfrac12\lsp(-1)^{\ell}\lsp F_{2,\lsp\Delta,\lsp\ell,\lsp\Delta_R}
^{\lsp\phib\phi;\lsp RR}\\
\vspace{4pt}\tfrac12\lsp(-1)^{\ell}\lsp F_{1,\lsp\Delta,\lsp\ell,
\lsp\Delta_R}^{\lsp\phib\phi;\lsp RR}& 0 & 0\\
\tfrac12\lsp(-1)^{\ell}\lsp F_{2,\lsp\Delta,\lsp\ell,\lsp\Delta_R}
^{\lsp\phib\phi;\lsp RR}& 0 & 0
\end{pmatrix}\,,\\
V^5_{\Delta,\lsp\ell,\lsp\Delta_R}=\begin{pmatrix}
0 & \tfrac12\lsp(-1)^{\ell+1}\lsp H_{1,\lsp\Delta,\lsp\ell,\lsp\Delta_R}
^{\phib\phi;\lsp RR} &
\tfrac12\lsp(-1)^{\ell+1}\lsp H_{2,\lsp\Delta,\lsp\ell,\lsp\Delta_R}
^{\phib\phi;\lsp RR}\\
\vspace{4pt}\tfrac12\lsp(-1)^{\ell+1}\lsp H_{1,\lsp\Delta,\lsp\ell,
\lsp\Delta_R}^{\phib\phi;\lsp RR} & 0 & 0\\
\tfrac12\lsp(-1)^{\ell+1}\lsp H_{2,\lsp\Delta,\lsp\ell
,\lsp\Delta_R}^{\phib\phi;\lsp RR} & 0 & 0
\end{pmatrix}\,,\\
V^6=\begin{pmatrix}
0 & 0 & 0\\
0 & 0 & 0\\
0 & 0 & 0
\end{pmatrix}\,,\qquad
V^7_{\Delta,\lsp\ell,\lsp\Delta_R}=\begin{pmatrix}
0 & 0 & 0\\
0 & F_{1,\lsp\Delta,\lsp\ell,\lsp\Delta_R}^{\lsp RR\lsp;\lsp RR} & 0\\
0 & 0 & F_{2,\lsp\Delta,\lsp\ell,\lsp\Delta_R}^{\lsp RR\lsp;\lsp RR}\\
\end{pmatrix}\,,
\stepcounter{equation}\tag{\theequation}
\end{gather*}
and the remaining vectors are given by
\eqn{\vec{W}_{\Delta,\lsp\ell,\lsp\Delta_\phi}=\begin{pmatrix}
  F^{\lsp\phib\phib;\lsp \phi\phi}_{\Delta,\lsp\ell,\lsp \Delta_\phi}
  \vspace{5pt}\\
  -H^{\phib\phib;\lsp \phi\phi}_{\Delta,\lsp\ell,\lsp \Delta_\phi}
  \vspace{5pt}\\
  0\\
  \vdots
  \\
  0
\end{pmatrix}\,,\qquad
\vec{\bar{X}}_{\Delta,\lsp\ell,\lsp\Delta_\phi,\lsp\Delta_R}=\begin{pmatrix}
0\\
0\\
0\vspace{5pt}\\
(-1)^\ell\lsp\bar{\CF}_{\Delta,\lsp\ell,\lsp\Delta_\phi,\lsp\Delta_R}^
{\lsp\phib R\lsp;\lsp R\phi}\vspace{5pt}\\
(-1)^\ell\lsp\bar{\CH}_{\Delta,\lsp\ell,\lsp\Delta_\phi,\lsp\Delta_R}^
{\phib R\lsp;\lsp R\phi}\vspace{5pt}\\
\bar{\CF}_{\lsp\Delta,\lsp\ell,\lsp
\Delta_\phi,\lsp\Delta_R}^{\lsp\phib R\lsp;\lsp \phi R}\vspace{5pt}\\
0
\end{pmatrix}\,,}[]
with definitions for $\vec{\hat{X}}$ and $\vec{\check{X}}$ similar to that
for $\vec{\bar{X}}$ but involving $\hat{\CF}, \hat{\CH}, \check{\CF},
\check{\CH}$, and
\eqn{\vec{Y}_{\Delta,\lsp\ell,\lsp\Delta_\phi,\lsp\Delta_R}=
\begin{pmatrix}
0\\
0\\
0\vspace{5pt}\\
(-1)^\ell\lsp F_{\lsp\Delta,\lsp\ell,\lsp\Delta_\phi,\lsp\Delta_R}^{\lsp
\phib R\lsp;\lsp R\phi}\vspace{5pt}\\
(-1)^\ell\lsp H_{\lsp\Delta,\lsp\ell,\lsp\Delta_\phi,\lsp\Delta_R}^{
\phib R\lsp;\lsp R\phi}\vspace{5pt}\\
F_{\lsp\Delta,\lsp\ell,\lsp
\Delta_\phi,\lsp\Delta_R}^{\lsp\phib R\lsp;\lsp \phi R}\vspace{5pt}\\
0
\end{pmatrix}\,,\qquad
\vec{Z}_{\Delta,\lsp\ell,\lsp\Delta_R}=\begin{pmatrix}
0\\
\vdots
\\
0\\
F_{\Delta,\lsp\ell,\lsp\Delta_R}^{\lsp RR;\lsp RR}
\end{pmatrix}\,.}[]

We should note here that the entries of $\vec{V}_{\Delta,\lsp \ell,\lsp
\Delta_\phi,\lsp \Delta_R}$ are $3\times3$ matrices because \evBl, \oddBl,
\RRBlockEven, and \RRBlockOdd do not contain their conformal block
contributions with fixed relative coefficients. The subscripts 1 and 2 in
the functions $F$ and $H$ of $V^4_{\Delta,\lsp \ell,\lsp \Delta_R},
V^5_{\Delta,\lsp \ell,\lsp \Delta_R}$ and $V^7_{\Delta,\lsp \ell,\lsp
\Delta_R}$ denote the first and second part of the corresponding $\CF$ and
$\CH$ functions defined in \rescBlI, \rescBlII and \FRR, as obtained when
the blocks \evBl, \oddBl, \RRBlockEven and \RRBlockOdd are used and the
coefficient $c_{RR\CO_\ell}^{\lsp\prime}$ is appropriately defined.  For
example, for even $\ell$ we have
$F_{2,\lsp\Delta,\lsp\ell,\lsp\Delta_R}^{\lsp\phib\phi;\lsp
RR}=-\frac{1}{16\Delta(\Delta-\ell-1)(\Delta+\ell+1)}\,g_{\Delta+2,
\lsp\ell}$ and $c_{RR\CO_{\ell}}^{\lsp\prime}
=(\Delta+\ell)^2c_{RR\CO_\ell}^{\lsp(0)} -8(\Delta-1)c_{RR\CO_\ell}^{\lsp
(2)}$ as follows from \evBl. Note that we can neglect
$\vec{Z}_{\Delta,\lsp\ell,\lsp \Delta_R}$ for its contributions are already
contained in $V^7_{\Delta,\lsp \ell, \lsp\Delta_R}$.

The crossing relation \CrossVec can be used with the usual numerical
methods. This requires polynomial approximations for derivatives of the
various functions that participate. We describe the required results in
Appendix~\ref{appApprox}. For numerical optimization we use
\texttt{SDPB}~\cite{Simmons-Duffin:2015qma}. The functional search space is
governed by the parameter $\Lambda$, where each component  $\alpha_i$ of a
seven-functional $\vec{\alpha}$ is a linear combination of
$\frac12\left\lfloor \frac{\Lambda+2}{2}\right\rfloor \left(\left\lfloor
\frac{\Lambda+2}{2} \right\rfloor + 1\right) $ independent nonvanishing
derivatives, $\alpha_i \propto \sum_{m,n} a_{mn}^i \partial_z^m
\partial_{\bar{z}}^n \big|_{1/2,1/2}$ with $m+n \leq \Lambda$. For example,
for $\Lambda=17$, a common choice in the plots below, the search space is
315-dimensional.

\newsec{Crossing relations with linear multiplets}[crossRelJ]
The crossing relations obtained in this case can be brought to the form
\eqna{&\sum_{\substack{\CO_\ell\in\phib\times\phi\\\CO_\ell\in J\times J}}
\hspace{-4pt}\begin{pmatrix}
  c^\ast_{\phib\phi\CO_\ell} & c^\ast_{JJ\CO_\ell}
\end{pmatrix}
\vec{V}_{\Delta,\lsp\ell,\lsp\Delta_\phi}
\begin{pmatrix}
c_{\phib\phi\CO_\ell}\\ c_{JJ\CO_\ell}
\end{pmatrix}\,\,
+\hspace{-4pt}\sum_{\COb_\ell\in\phib\times\phib}\hspace{-4pt}|c_{\phib\phib\CO_\ell}|^2
\vec{W}_{\Delta,\lsp\ell,\lsp\Delta_\phi}\\
&\hspace{0.7cm}+\hspace{-4pt}\sum_{\COb\in\phib\times J}\hspace{-4pt}
|\bar{c}_{\phib J\CO}|^2\vec{\bar{X}}_{\Delta,\lsp0,
\lsp\Delta_\phi}
+\hspace{-4pt}\sum_{(Q\COb)_\ell\in\phib\times J}\hspace{-4pt}
|\hat{c}_{\phib J(\Qb\CO)_\ell}|^2
\vec{\hat{X}}_{\Delta,\lsp\ell,\lsp\Delta_\phi}
+\hspace{-4pt}\sum_{(Q\COb)_\ell\in\phib\times J}\hspace{-4pt}
|\check{c}_{\phib J(\Qb\CO)_\ell}|^2
\vec{\check{X}}_{\Delta,\lsp\ell,\lsp\Delta_\phi}\\
&\hspace{5.6cm}+\hspace{-8pt}\sum_{(Q^2\COb)\in\phib\times J}\hspace{-8pt}
|c_{\phib J(\Qb^2\CO)}|^2\lsp\vec{Y}_{\Delta,\lsp\ell,\lsp\Delta_\phi}
+\hspace{-10pt}\sum_{(Q\CO)_\ell\in J\times J}\hspace{-8pt}
|c_{JJ(Q\CO)_\ell}|^2\vec{Z}_{\Delta,\lsp\ell}=0\,,}[CrossVecJ]
where $\vec{\bar{X}}_{\Delta,\lsp0,\lsp\Delta_\phi}$ goes over just two
scalar operators with dimension $\Delta_\phi$ and $\Delta_\phi+2$.  Due to
the determined coefficients in the superconformal blocks \evJBl, \oddJBl,
\JJBlockEven, and \JJBlockOdd, the seven-vector
$\vec{V}_{\Delta,\lsp\ell,\lsp\Delta_\phi}$ contains $2\times 2$ matrices
now, contrary to the case in \CrossVec where
$\vec{V}_{\Delta,\lsp\ell,\lsp\Delta_\phi,\lsp\Delta_R}$ contained $3\times
3$ matrices. Here, $\vec{V}_{\Delta,\lsp\ell,\lsp\Delta_\phi}$ contains the
matrices
\eqn{\begin{gathered}
V^1_{\Delta,\lsp\ell,\lsp\Delta_\phi}=\begin{pmatrix}
\CF^{\lsp\phib\phi;\lsp
\phi\phib}_{\Delta,\lsp\ell,\lsp\Delta_\phi} & 0\\
0 & 0
\end{pmatrix}\,,\qquad
V^2_{\Delta,\lsp\ell,\lsp\Delta_\phi}=\begin{pmatrix}
\CH^{\phib\phi;\lsp
\phi\phib}_{\Delta,\lsp\ell,\lsp\Delta_\phi} & 0 \\
0 & 0
\end{pmatrix}\,,\\
V^3_{\Delta,\lsp\ell,\lsp\Delta_\phi}=\begin{pmatrix}
(-1)^\ell\CF^{\lsp\phib\phi;\lsp
\phib\phi}_{\Delta,\lsp\ell,\lsp\Delta_\phi} & 0\\
0 & 0
\end{pmatrix}\,,\qquad
V^4_{\Delta,\lsp\ell}=\begin{pmatrix}
  0 & \tfrac12\lsp(-1)^{\ell}\lsp \CF_{\Delta,\lsp\ell}^
  {\lsp\phib\phi;\lsp JJ}\\
\vspace{4pt}\tfrac12\lsp(-1)^{\ell}\lsp \CF_{\Delta,\lsp\ell}
^{\lsp\phib\phi;\lsp JJ} & 0
\end{pmatrix}\,,\\
V^5_{\Delta,\lsp\ell}=\begin{pmatrix}
0 & \tfrac12\lsp(-1)^{\ell+1}\lsp \CH_{\Delta,\lsp\ell}
^{\phib\phi;\lsp JJ}\\
\vspace{4pt}\tfrac12\lsp(-1)^{\ell+1}\lsp
\CH_{\Delta,\lsp\ell}^{\phib\phi;\lsp JJ} & 0
\end{pmatrix}\,,\quad
V^6=\begin{pmatrix}
0 & 0\\
0 & 0
\end{pmatrix}\,,\quad
V^7_{\Delta,\lsp\ell}=\begin{pmatrix}
0 & 0\\
0 & \CF_{\Delta,\lsp\ell}^{\lsp JJ;\lsp JJ}
\end{pmatrix}\,,
\end{gathered}}[]
and the remaining vectors are given by
\eqn{\vec{W}_{\Delta,\lsp\ell,\lsp\Delta_\phi}=\begin{pmatrix}
  F^{\lsp\phib\phib;\lsp \phi\phi}_{\Delta,\lsp\ell,\lsp \Delta_\phi}
  \vspace{5pt}\\
  -H^{\phib\phib;\lsp \phi\phi}_{\Delta,\lsp\ell,\lsp \Delta_\phi}
  \vspace{5pt}\\
  0\\
  \vdots
  \\
  0
\end{pmatrix}\,,\qquad
\vec{\bar{X}}_{\Delta,\lsp\ell,\lsp\Delta_\phi}=\begin{pmatrix}
0\\
0\\
0\vspace{5pt}\\
\bar{F}_{\Delta,\lsp0,\lsp\Delta_\phi}^{\lsp\phib
J;\lsp J\phi}\vspace{5pt}\\
\bar{H}_{\Delta,\lsp0,\lsp\Delta_\phi}^{\phib
J;\lsp J\phi}\vspace{5pt}\\
\bar{F}_{\lsp\Delta,\lsp0,\lsp
\Delta_\phi}^{\lsp\phib J;\lsp \phi J}\vspace{5pt}\\
0
\end{pmatrix}\,,\qquad
\vec{\hat{X}}_{\Delta,\lsp\ell,\lsp\Delta_\phi}=\begin{pmatrix}
0\\
0\\
0\vspace{5pt}\\
(-1)^\ell\lsp\hat{\CF}_{\Delta,\lsp\ell,\lsp\Delta_\phi}^{\lsp\phib
J;\lsp J\phi}\vspace{5pt}\\
(-1)^\ell\lsp\hat{\CH}_{\Delta,\lsp\ell,\lsp\Delta_\phi}^{\phib
J;\lsp J\phi}\vspace{5pt}\\
\hat{\CF}_{\lsp\Delta,\lsp\ell,\lsp
\Delta_\phi}^{\lsp\phib J;\lsp \phi J}\vspace{5pt}\\
0
\end{pmatrix}
\,,}[]
with a similar definition for $\vec{\check{X}}$, and
\eqn{\vec{Y}_{\Delta,\lsp\ell,\lsp\Delta_\phi}=
\begin{pmatrix}
0\\
0\\
0\vspace{5pt}\\
(-1)^\ell\lsp F_{\lsp\Delta,\lsp\ell,\lsp\Delta_\phi}^{\lsp
\phib J;\lsp J\phi}\vspace{5pt}\\
(-1)^\ell\lsp H_{\lsp\Delta,\lsp\ell,\lsp\Delta_\phi}^{
\phib J;\lsp J\phi}\vspace{5pt}\\
F_{\lsp\Delta,\lsp\ell,\lsp
\Delta_\phi}^{\lsp\phib J;\lsp \phi J}\vspace{5pt}\\
0
\end{pmatrix}\,,\qquad
\vec{Z}_{\Delta,\lsp\ell}=\begin{pmatrix}
0\\
\vdots
\\
0\\
F_{\Delta,\lsp\ell}^{\lsp JJ;\lsp JJ}
\end{pmatrix}\,.}[]
The various functions $F, \CF$ and $H, \CH$ here are defined similarly to
the analogous functions defined in section \crossRel, using the
superconformal blocks of section \secJBl. We note that contrary to the case
in section \crossRel, the contributions of $\vec{Z}_{\Delta,\lsp\ell}$ are
not identical to those in $V^7_{\Delta,\lsp\ell}$, and so
$\vec{Z}_{\Delta,\lsp\ell}$ needs to be included in our numerical analysis.

\newsec{Bounds in theories with \texorpdfstring{$\phi$}{phi} and
\texorpdfstring{$R$}{R}}[BoundR]

\begin{figure}[H]
  \begin{center}
    \includegraphics{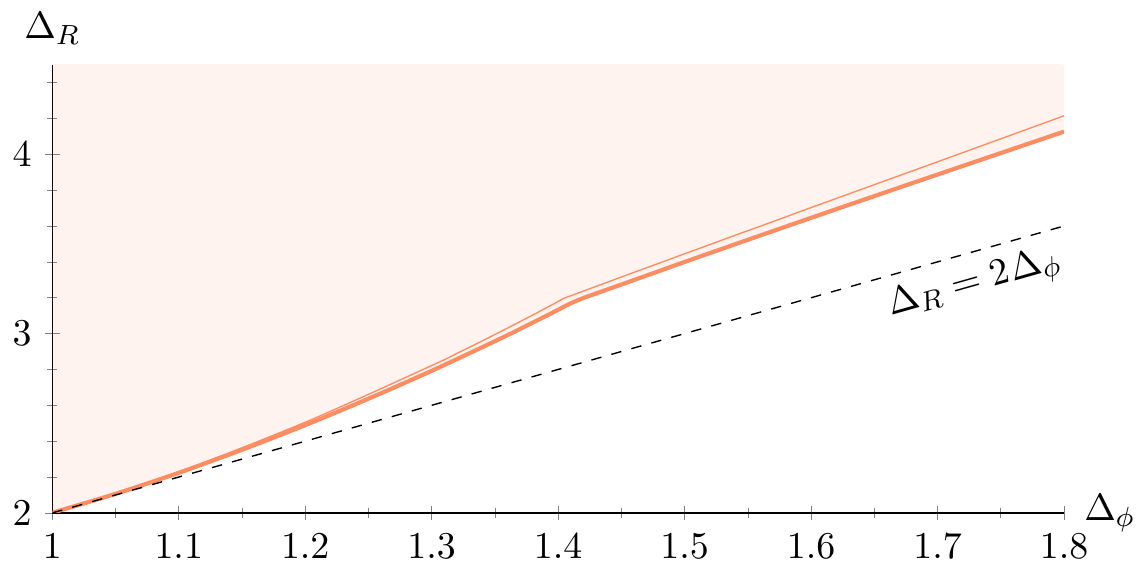}
  \end{center}
  \vspace{-11pt}
  \caption{Upper bound on the dimension of the operator $R$ as a function
  of $\Delta_\phi$ using only \crossRelphi. The generalized free theory
  dashed line $\Delta_R=2\hspace{0.5pt}\Delta_{\phi}$ is also shown.  The
  shaded area is excluded. In this plot we use $\Lambda=21$ for the thin
  and $\Lambda=29$ for the thick line.}
  \label{fig:dim_phibphi}
\end{figure}

\subsec{Using only the chiral-chiral and chiral-antichiral crossing
relations}
A bound on the dimension of the first unprotected scalar operator $R$ in
the $\phib\times\phi$ OPE using just \crossRelphi was first obtained
in~\cite{Poland:2011ey} and recently reproduced in~\cite{Poland:2015mta}.
This bound, for $\Lambda=21$ and $\Lambda=29$, is shown in
Fig.~\ref{fig:dim_phibphi}, and displays a mild kink at
$\Delta_\phi\approx1.4$. The bound for $\Lambda=21$ was first obtained
in~\cite{Poland:2011ey}. Here we provide a slightly stronger bound at
$\Lambda=29$.
If we assume that $\phi^2=0$, then the allowed region on the left of the
kink disappears~\cite{Bobev:2015jxa, Poland:2015mta}, turning the kink into
a sharp corner. The precision analysis of \cite{Poland:2015mta} suggests
that the kink is at $\Delta_\phi=\frac{10}{7}$, although this relies on
extrapolation.

\begin{figure}[H]
  \begin{center}
    \includegraphics{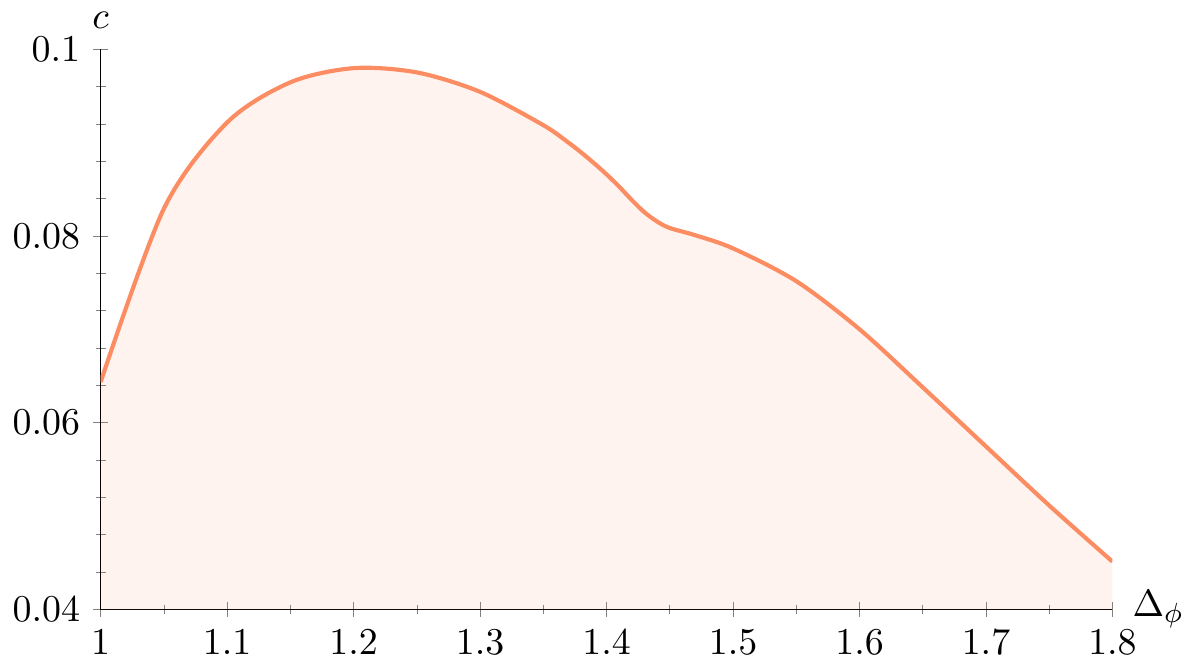}
  \end{center}
  \vspace{-11pt}
  \caption{Lower bound on the central charge as a function of
  $\Delta_\phi$.  The shaded area is excluded. In this plot we use
  $\Lambda=25$.}
  \label{fig:cc}
\end{figure}

\begin{figure}[H]
  \begin{center}
    \includegraphics{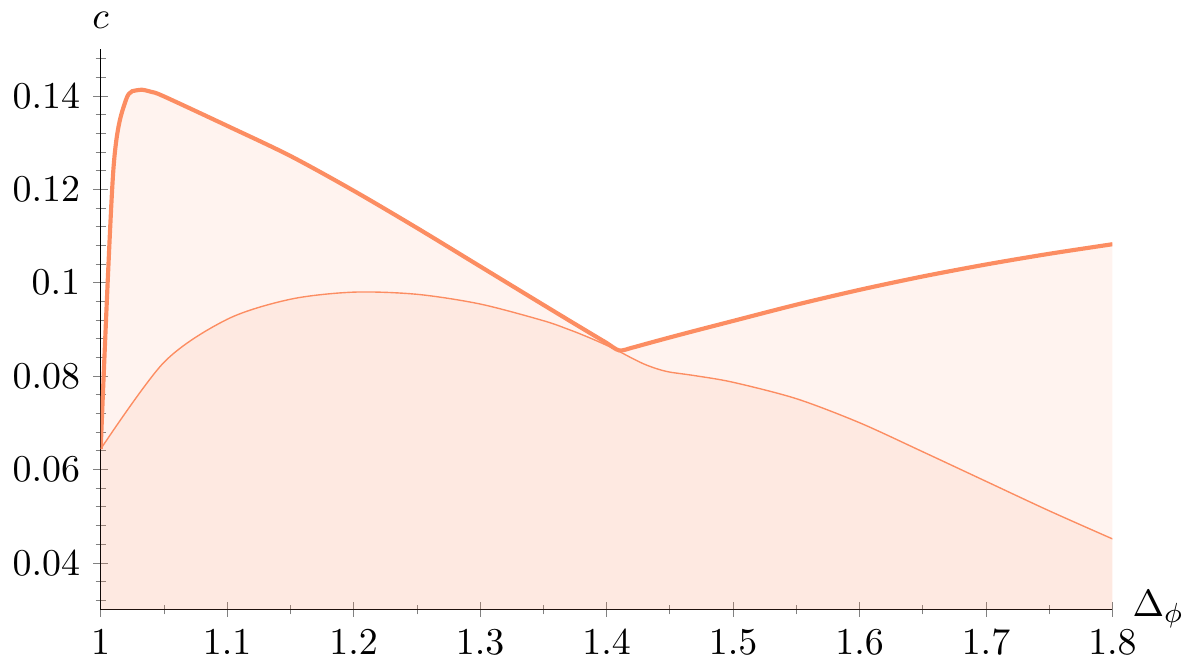}
  \end{center}
  \vspace{-11pt}
  \caption{The thick line is the lower bound on the central charge as a
  function of $\Delta_\phi$, assuming that $\Delta_R$ lies on the bound of
  Fig.~\ref{fig:dim_phibphi}. The thin line is the bound of
  Fig.~\ref{fig:cc}. The shaded area is excluded. In this plot we use
  $\Lambda=25$.}
  \label{fig:cc_assum}
\end{figure}

Using \crossRelphi we can also obtain a lower bound on the central charge.
This is shown in Fig.~\ref{fig:cc} for $\Lambda=25$. The corresponding
bound for $\Lambda=21$ first appeared in~\cite{Poland:2011ey}, and was
later improved in~\cite{Poland:2015mta}. The bound contains a feature
slightly to the right of the kink of Fig.~\ref{fig:dim_phibphi}. Close to
the origin the bound sharply falls just below the free chiral multiplet
value of $c=\frac{1}{24}$ in our normalization~\cite{Poland:2010wg}.
We may further assume that $\Delta_R$ lies on the bound of
Fig.~\ref{fig:dim_phibphi}, and that $R$ is the first scalar after the
identity operator in the $\phib\times\phi$ OPE. The lower bound on the
central charge obtained in this case is shown in Fig.~\ref{fig:cc_assum}.
As we see, these extra assumptions strengthen the bound globally, but have
the weakest effect around the free theory and $\Delta_\phi\approx1.4$. At
that $\Delta_\phi$, which coincides with the position of the kink, we
observe a local minimum of the lower bound on $c$. This result has also
been discussed in~\cite{Bobev:2015jxa}, and is similar to the corresponding
bound obtained in $d=3$ in~\cite{ElShowk:2012ht}, although the free theory
of a single chiral operator in our case has a lower $c$ than the minimum in
Fig.~\ref{fig:cc_assum}. The assumption $\phi^2=0$ excludes the region to
the left of $\Delta_\phi\approx1.4$. Therefore, we may conjecture that the
putative theory that lives on the kink minimizes $c$ among $\CN=1$
superconformal theories that have a chiral operator $\phi$ that satisfies
$\phi^2=0$. Such theories were obtained recently~\cite{Xie:2016hny,
Buican:2016hnq} from deformations of $\CN=2$ Argyres--Douglas
theories~\cite{Argyres:1995jj, Argyres:1995xn, Eguchi:1996vu}, but they
appear to have larger $c$ than the one obtained for the minimal theory
in~\cite{Poland:2015mta}, namely $c_{\text{minimal}}=\frac19$ after
extrapolating to $\Lambda\rightarrow\infty$.
\subsec{Using the full set of crossing relations involving
  \texorpdfstring{$\phi$}{phi} and \texorpdfstring{$R$}{R}}[subsecFSR]
We will now explore bootstrap constraints using the full system of crossing
relations \CrossVec. The virtue of considering mixed correlators is that
they allow us to probe a larger part of the operator spectrum, e.g.\ we can
obtain bounds on operator dimensions and OPE coefficients of operators in
the $\phib\times R$ OPE. In this subsection we assume that $\Delta_R$ lies
on the (stronger) bound of Fig.~\ref{fig:dim_phibphi}. We also impose
$c_{\phib R\phi}=c_{\phib\phi R}$---the implementation of this
follows~\cite{Kos:2015mba}, i.e.\ we add a single constraint for
$\vec{V}_{\Delta_R,\lsp0, \lsp\Delta_\phi,\lsp\Delta_R} +
\vec{\bar{X}}_{\Delta_\phi,\lsp0,\lsp \Delta_\phi,\lsp \Delta_R}
\otimes\text{diag}(1,0,0)$ to our optimization problem. Finally, we
introduce a gap of one between the dimension of $R$ and that of the next
unprotected real scalar in the spectrum, $R^\prime$. We have found that for
low values of this gap the bounds below are not sensitive to the choice of
the gap.

\begin{figure}[ht]
  \begin{center}
    \includegraphics{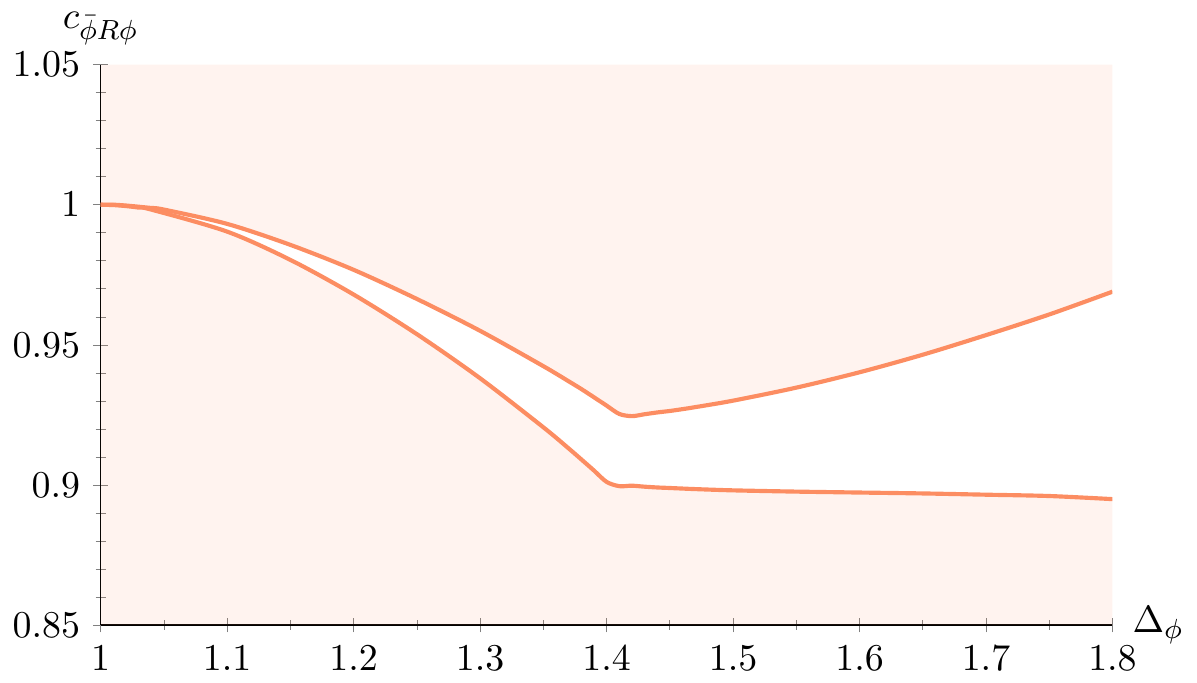}
  \end{center}
  \vspace{-11pt}
  \caption{Upper and lower bounds on the OPE coefficient of the operator
  $\bar{\phi}$ in the $\bar{\phi}\times R$ OPE as a function of
  $\Delta_\phi$, assuming $\Delta_R$ lies on the bound of
  Fig.~\ref{fig:dim_phibphi} and demanding
  $c_{\bar{\phi}R\phi}=c_{\bar{\phi}\phi R}$. We also impose a gap equal to
  one between $\Delta_R$ and $\Delta_{R^\prime}$. The shaded area is
  excluded.  In this plot we use $\Lambda=17$.}
  \label{fig:phi_ope_coeff_R_at_bound}
\end{figure}

First we would like to obtain a bound on the OPE coefficient of the
operator $\phib$ in the $\phib\times R$ OPE. We can obtain both an upper
and a lower bound; they are both shown in
Fig.~\ref{fig:phi_ope_coeff_R_at_bound}.
As we see there is a minimum of the upper bound slightly to the right of
$\Delta_\phi\approx1.4$. Note that the bound of $c_{\phib R\phi}$ at the
minimum is lower than the free theory value which is equal to one.

\begin{figure}[ht]
  \begin{center}
    \includegraphics{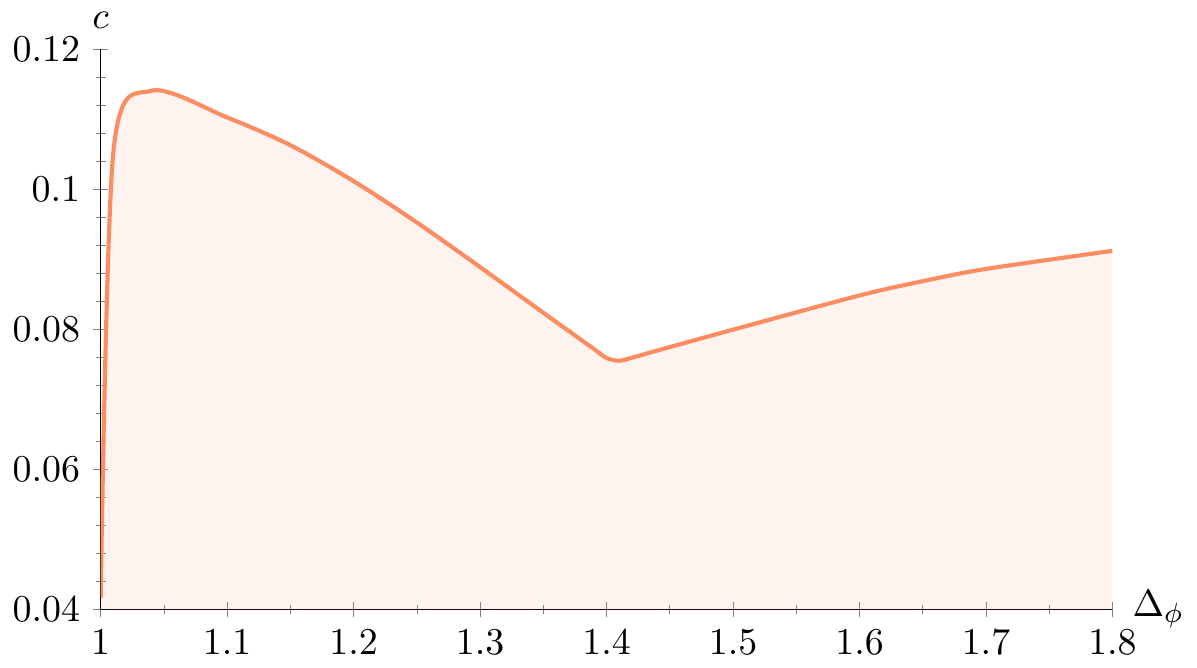}
  \end{center}
  \vspace{-11pt}
  \caption{Lower bound on the central charge as a function of
  $\Delta_\phi$, assuming that $\Delta_R$ lies on the bound of
  Fig.~\ref{fig:dim_phibphi} and demanding $c_{\phib R\phi}=c_{\phib\phi
  R}$. We also impose a gap equal to one between $\Delta_R$ and
  $\Delta_{R^\prime}$. The shaded area is excluded. In this plot we use
  $\Lambda=17$.}
  \label{fig:cc_assum_mixed}
\end{figure}

Using mixed correlators we can also obtain a bound on the central charge
similar to that of Fig.~\ref{fig:cc_assum}, i.e.\ assuming that $\Delta_R$
saturates its bound. The bound is shown in Fig.~\ref{fig:cc_assum_mixed}.
As we see, even though we use the mixed correlator crossing relations the
bound obtained is very similar to the corresponding bound in
Fig.~\ref{fig:cc_assum}. The bound of Fig.~\ref{fig:cc_assum_mixed} is
weaker than that of Fig.~\ref{fig:cc_assum} due to the lower $\Lambda$ used
in the former.

With the inclusion of the crossing relations \crossRelphiRI,
\crossRelphiRII and \crossRelphiRIII we can attempt to constrain scaling
dimensions of operators with R-charge equal to that of $\phib$. In
particular, we can attempt to find a bound on the dimension of the first
scalar superconformal primary after $\phib$ in the $\phib\times R$ OPE,
called $\phib^{\lsp\prime}$, assuming that $\Delta_R$ lies on the
(stronger) bound of Fig.~\ref{fig:dim_phibphi}.

\begin{figure}[H]
  \begin{center}
    \includegraphics{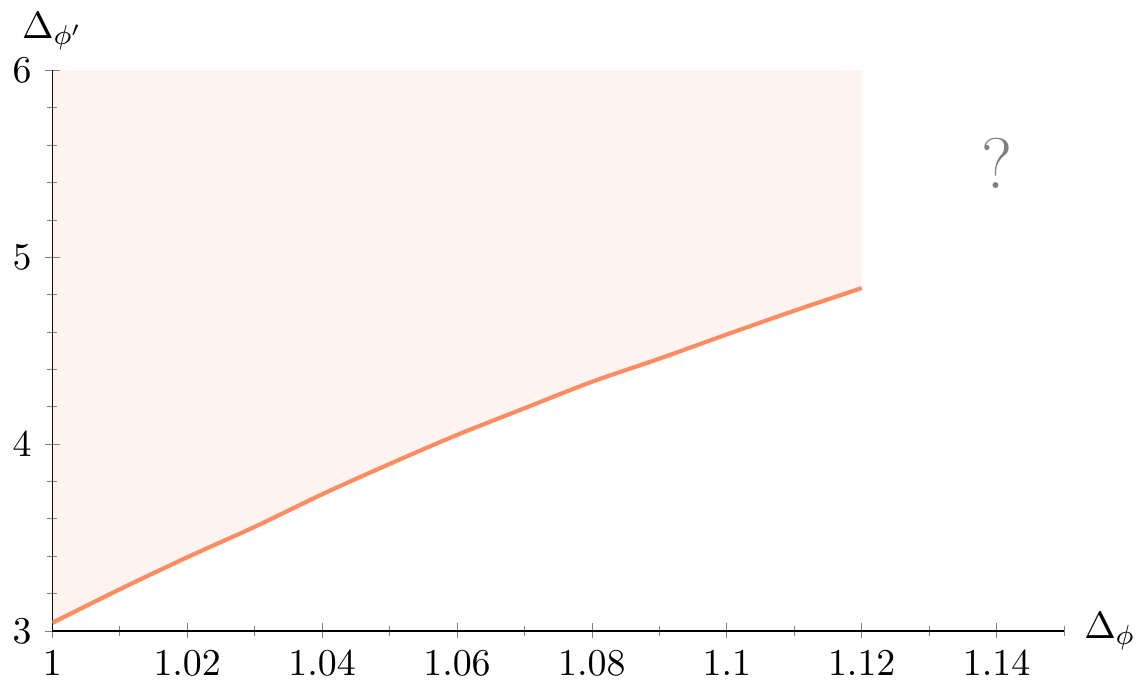}
  \end{center}
  \vspace{-11pt}
  \caption{Upper bound on $\Delta_{\phi^\prime}$ as a function of
  $\Delta_\phi$, assuming that $\Delta_R$ lies on the bound of
  Fig.~\ref{fig:dim_phibphi} and imposing $c_{\phib R\phi}=c_{\phib\phi
  R}$. Here we neglect $\vec{\hat{X}}$ and $\vec{Y}$ scalar contributions
  in \CrossVec, and impose a gap equal to one between $\Delta_R$ and
  $\Delta_{R^\prime}$. The shaded area is excluded. In this plot we use
  $\Lambda=11$.}
  \label{fig:phip}
\end{figure}

Numerically, this turned out to be a hard problem. For $\Lambda=11$ a bound
on $\Delta_{\phi^{\prime}}$ did not arise for any value of $\Delta_{\phi}$.
With the assumption that there are no $Q$-exact scalar operators in the
$\phib\times R$ OPE, i.e.\ neglecting the $\vec{\hat{X}}$ and $\vec{Y}$
scalar contributions in \CrossVec, we managed to obtain a bound on
$\Delta_{\phi^{\prime}}$ but only for $\Delta_{\phi}\lesssim1.12$, after
which point the bound was abruptly lost. This bound is shown in
Fig.~\ref{fig:phip}.
Increasing our functional search space by taking $\Lambda=13$, $\Lambda=17$
and $\Lambda=19$ we find a bound on $\Delta_{\phi^{\prime}}$ up to
$\Delta_\phi\approx1.27$, $\Delta_\phi\approx1.32$ and
$\Delta_\phi\approx1.34$, respectively.  At the corresponding $\Delta_\phi$
the bound is again abruptly lost. Note that for these results we do not
actually obtain the bound, but rather we ask if the spectrum with $\phib$
as the only scalar in the $\phib\times R$ OPE is allowed or not.  We
believe that numerical analysis for higher $\Lambda$ will yield bounds on
$\Delta_{\phi^{\prime}}$ for higher $\Delta_\phi$, but it is puzzling that
in going from $\Lambda=17$ to $\Lambda=19$ we have a very small gain in the
$\Delta_\phi$ up to which a bound on $\Delta_{\phi^\prime}$ can be
obtained.

The various features we have seen in plots of this section indicate the
existence of a CFT with a chiral operator of dimension
$\Delta_{\phi}\approx1.4$, or $\Delta_\phi=\frac{10}{7}$ based on the
analysis of~\cite{Poland:2015mta}. Unfortunately the mixed correlator
analysis has not allowed us to isolate this putative CFT from the allowed
region around it, particularly from the allowed region for higher
$\Delta_\phi$. We remind the reader that the region for
$\Delta_\phi<\frac{10}{7}$ can be excluded by imposing that $\phi^2=0$ as a
primary~\cite{Bobev:2015jxa, Poland:2015mta}. The set of conditions that
isolate this putative CFT from solutions to crossing symmetry with higher
$\Delta_\phi$ have not been found in this paper. We hope that future work
will be able to identify these conditions, or uncover a physical reason for
their absence.

\newsec{Bounds in theories with global symmetries}[BoundJ]

\subsec{Using the crossing relation from \texorpdfstring{$\langle JJJJ\rangle$}{<JJJJ>}}
Bootstrap bounds arising from the four-point function $\vev{J(x_1)\lsp
J(x_2)\lsp J(x_3)\lsp J(x_4)}$ were obtained recently
in~\cite{Berkooz:2014yda}. In fact,~\cite{Berkooz:2014yda} considered the
more complicated nonabelian case. Here we will consider just the Abelian
case, where $J$ carries no adjoint index, and obtain some further bounds
that have not appeared before.

Since the dimension of $J$ is fixed by symmetry, no external
operator dimension can be used as a free parameter. For the
plots in this section we will instead use the dimension of the first
unprotected operator $\CO$ in the $J\times J$ OPE as the parameter in the
horizontal axis. Note that there is an upper bound to how large that
dimension can get, and so our plots will not extend past that bound. This
bound is found here by looking at the value for which the square of the
plotted OPE coefficient turns negative.

\begin{figure}[ht]
  \centering
  \includegraphics{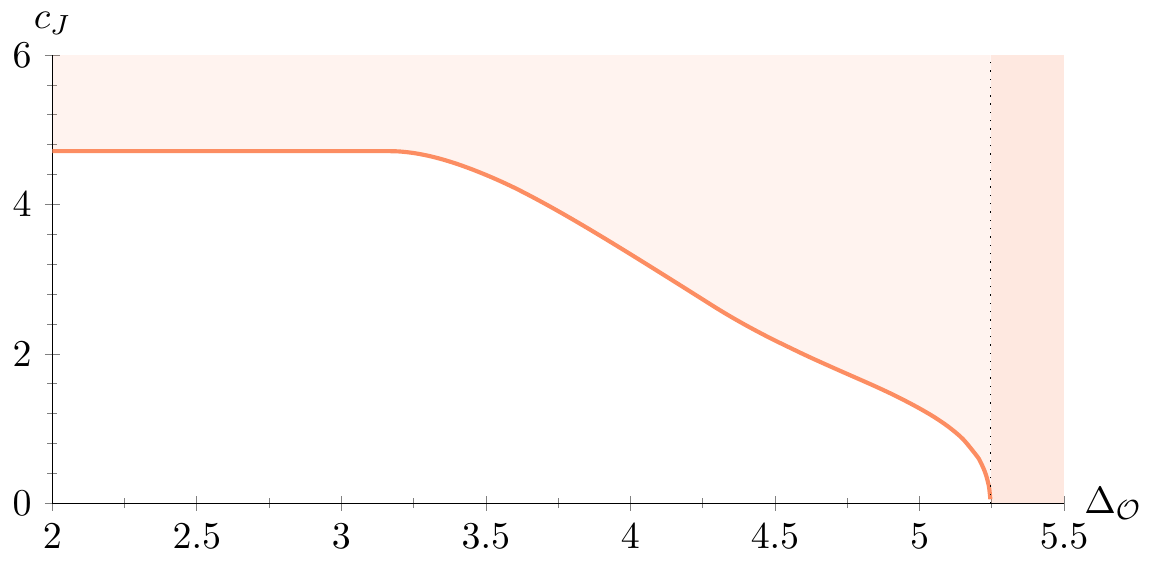}
  \caption{Upper bound on the OPE coefficient of $J$ in the $J\times J$ OPE
  as a function of the dimension of the first unprotected scalar in the
  $J\times J$ OPE. The region to the right of the dotted vertical line at
  $\Delta_{\mathcal{O}}=5.246$ is not allowed. In this plot we use
  $\Lambda=29$.}
  \label{fig:JCoeffBoundU1}
\end{figure}

\begin{figure}[H]
  \centering
  \includegraphics{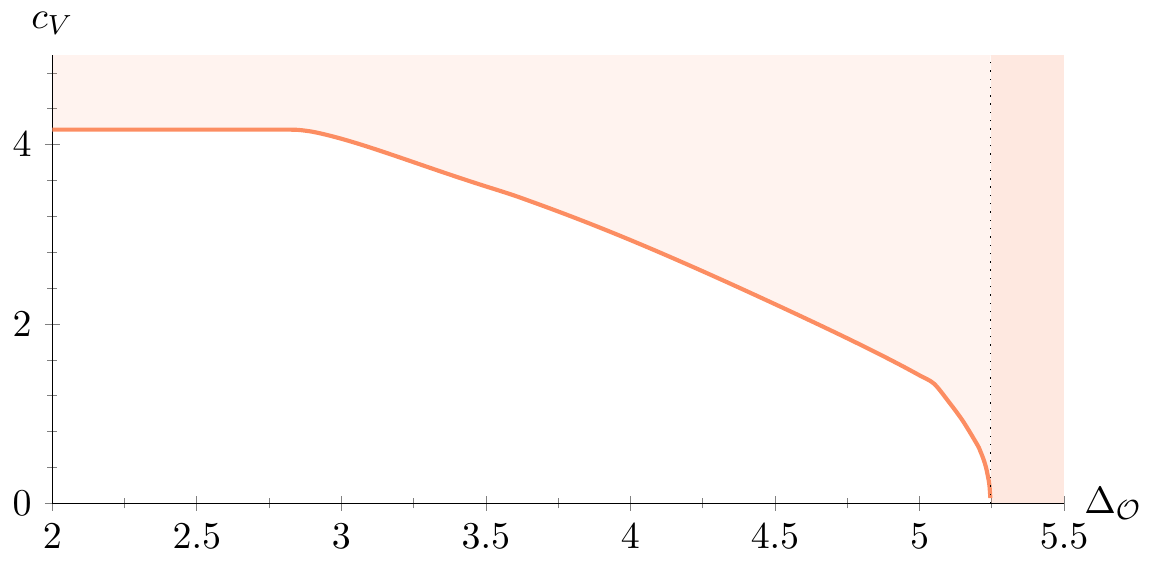}
  \caption{Upper bound on the OPE coefficient of the contributions to the
  $J\times J$ OPE arising from the leading vector superconformal primary
  $V$ as a function of the dimension of the first unprotected scalar in the
  $J\times J$ OPE. The region to the right of the dotted vertical line at
  $\Delta_{\mathcal{O}}=5.246$ is not allowed. In this plot we use
  $\Lambda=29$.}
  \label{fig:VCoeffBoundU1}
\end{figure}

First, we obtain an upper bound on the OPE coefficient of $J$ in the
$J\times J$ OPE. The bound is shown in Fig.~\ref{fig:JCoeffBoundU1}. It
contains a plateau that eventually breaks down, leading to a violation of
unitarity past $\Delta_\CO=5.246$. This is a reflection of the fact that
the dimension of the first unprotected scalar in the $J\times J$ OPE cannot
be larger than $\Delta_\CO=5.246$ consistently with unitarity.
\indent The $J\times J$ OPE also contains contributions arising from the
dimension-three vector multiplet that contains the stress-energy tensor. We
can obtain a bound on the OPE coefficient $c_V$ of these contributions; see
Fig.~\ref{fig:VCoeffBoundU1}. A lower bound on the central charge $c$ can
then be derived from these results, since $c_V^{\lsp 2}=\frac{1}{90\lsp c}$
in our conventions. Close to the origin we get $c\gtrsim 0.00064$, a bound
much weaker than that in Fig.~\ref{fig:cc}.

\begin{figure}[H]
  \begin{center}
    \includegraphics[width=.46\textwidth]{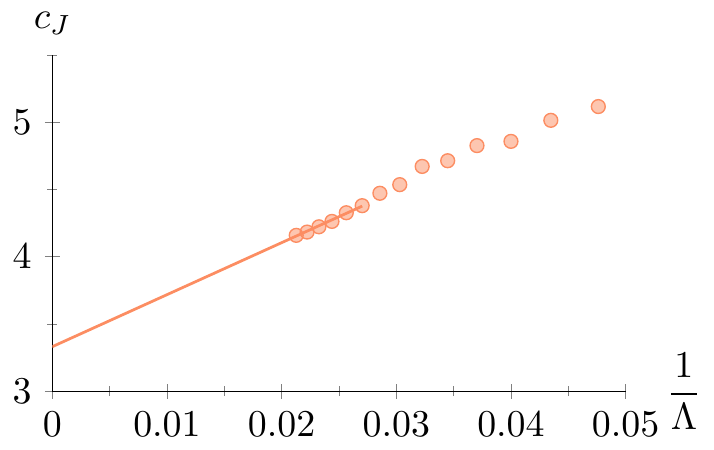}
    \hspace{.5cm}
    \includegraphics[width=.46\textwidth]{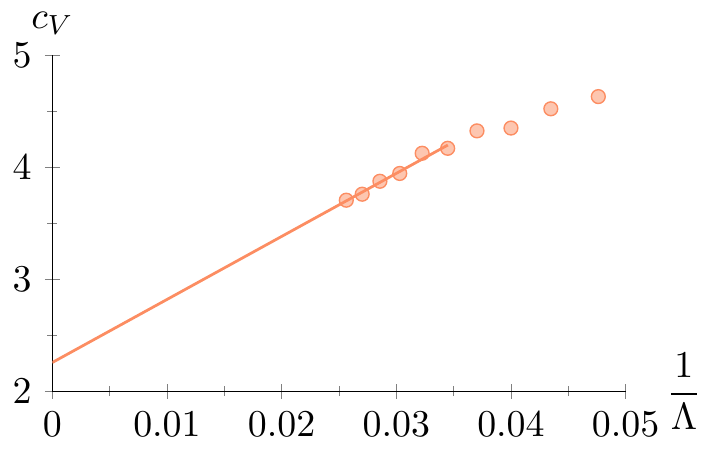}
  \end{center}
  \vspace{-12pt}
  \caption{The upper bounds on $c_J$ and $c_V$ with $\Delta_\CO=2$ as
  functions of the inverse cutoff $1/\Lambda$, and linear extrapolations of
  the six points closest to the origin.}
  \label{fig:cJcVExtrap}
\end{figure}

The bounds in Figs.~\ref{fig:JCoeffBoundU1} and~\ref{fig:VCoeffBoundU1}
were obtained using $\Lambda=29$.\foot{For lower values of $\Lambda$,
e.g.~$\Lambda=21$, we do not find an upper bound on $\Delta_\CO$, i.e.\
$c_J^2$ and $c_V^2$ never turn negative. The upper bounds for $c_J$ and
$c_V$ in those cases converge to values that do not change with
$\Delta_\CO$ no matter how large $\Delta_\CO$ becomes.}  We can also obtain
bounds for other values of $\Lambda$. We do this here letting $\CO$
saturate its unitarity bound, i.e.\ choosing $\Delta_\CO=2$. The plots are
shown in Fig.~\ref{fig:cJcVExtrap}. As $\Lambda$ gets larger we see observe
an approximately linear distribution of the bounds, which we then fit and
extrapolate to the origin. The fits are given by
\eqn{c_J^{\text{(fit)}}=3.311+\frac{39.412}{\Lambda}\,,\qquad
c_V^{\text{(fit)}}=2.256+\frac{56.279}{\Lambda}\,.
}[]
The limit $\Lambda\to\infty$ gives us an estimate of the converged optimal
bound that can be obtained.

\begin{figure}[H]
  \centering
  \includegraphics{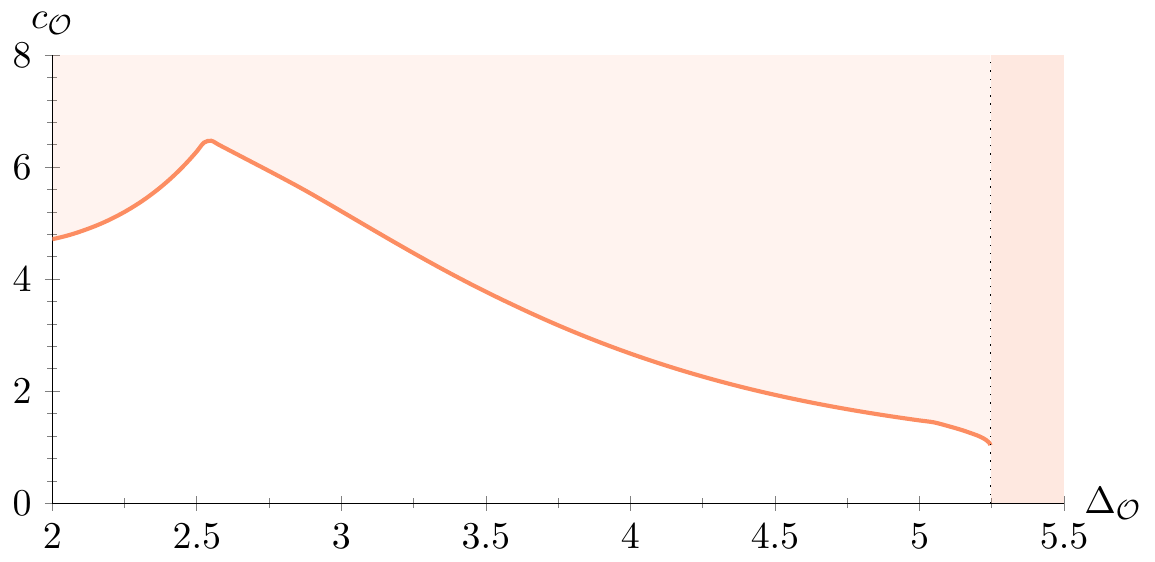}
  \caption{Upper bound on the OPE coefficient of the first unprotected
  scalar operator in the $J\times J$ OPE as a function of its dimension.
  The region to the right of the dotted vertical line at
  $\Delta_{\mathcal{O}}=5.246$ is not allowed.  In this plot we use
  $\Lambda=29$.}
  \label{fig:OCoeffBoundU1}
\end{figure}

Finally, we also find an upper bound on the OPE coefficient of $\CO$ as a
function of the dimension of $\CO$; see Fig.~\ref{fig:OCoeffBoundU1}.

\begin{figure}[H]
  \begin{center}
    \includegraphics{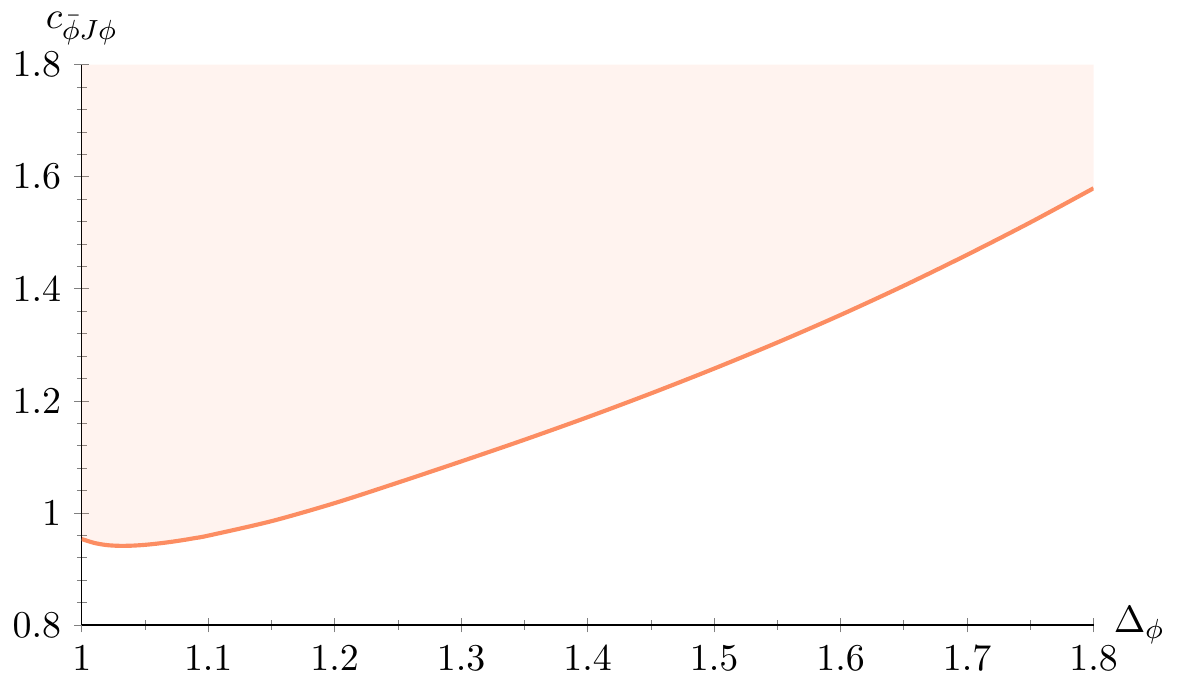}
  \end{center}
  \vspace{-11pt}
  \caption{Upper bound on the OPE coefficient of the operator $\bar{\phi}$
  in the $\bar{\phi}\times J$ OPE as a function of $\Delta_\phi$, demanding
  $c_{\phib J\phi}=c_{\phib\phi J}$. In this plot we use $\Lambda=17$.}
  \label{fig:phi_ope_coeff}
\end{figure}

\subsec{Using the full set of crossing relations involving
  \texorpdfstring{$\phi$}{phi} and \texorpdfstring{$J$}{J}}
Similarly to subsection \subsecFSR we can here obtain constraints on
operators that appear in the $\phib\times J$ OPE. One such operator is
$\phib$ itself, and we can obtain a bound on its OPE coefficient. This OPE
coefficient is equal to that of $J$ in the $\phib\times\phi$ OPE, and its
meaning has been analyzed in~\cite{Poland:2010wg}, where it was denoted by
$\tau_{IJ}T^I_{1\bar{1}}T^J_{1\bar{1}}$. The bound is shown in
Fig.~\ref{fig:phi_ope_coeff}.
One application of this bound is in $\text{SU}(N_c)$ SQCD with $N_f$
flavors $Q^i$ and $\tilde{Q}_{\tilde{\imath}}$. Mesons in this theory have
scaling dimension $\Delta_M=3(1-N_c/N_f)$, which can be close to one at the
lower end of the conformal window, $N_f\sim\frac32 N_c$. This was
considered first in~\cite{Poland:2010wg}, where the meson $M_1^1$ was taken
as the chiral operator and the relation
\eqn{\tau_{IJ}T_{1\bar{1}}^IT_{1\bar{1}}^J=2\frac{N_f-1}{3N_c{\!}^2}}[]
was obtained for the contributions of the flavor currents of the symmetry
group $\text{SU}(N_f)_\text{L}\times\text{SU}(N_f)_\text{R}$ of SQCD. This
satisfies our bound in Fig.~\ref{fig:phi_ope_coeff} comfortably. For
example, for $N_c=3$ and $N_f=5$, in which case $\Delta_M=1.2$, we have
$\tau_{IJ}T_{1\bar{1}}^IT_{1\bar{1}}^J\approx0.3$ with the bound
constraining this to be lower than approximately one. Even with these
numerical results we are far away from saturating the bound with SQCD,
although we can hope that by pushing the numerics further we will get much
closer in the near future.

We should also note here that very close to $\Delta_\phi=1$ our bound
appears to be converging to a value for $c_{\phib J\phi}$ below one, thus
excluding the free theory of a free chiral operator charged under a
$\text{U}(1)$.  While we have not been able to obtain a bound very close to
one, i.e.\ $10^{-15}$ or so away from it, we believe that the bound
abruptly jumps right above one as $\Delta_\phi\to1$ in order to allow the
free theory solution. This behavior of the bound has also been seen
in~\cite{Poland:2011ey}.

\begin{figure}[H]
  \begin{center}
    \includegraphics{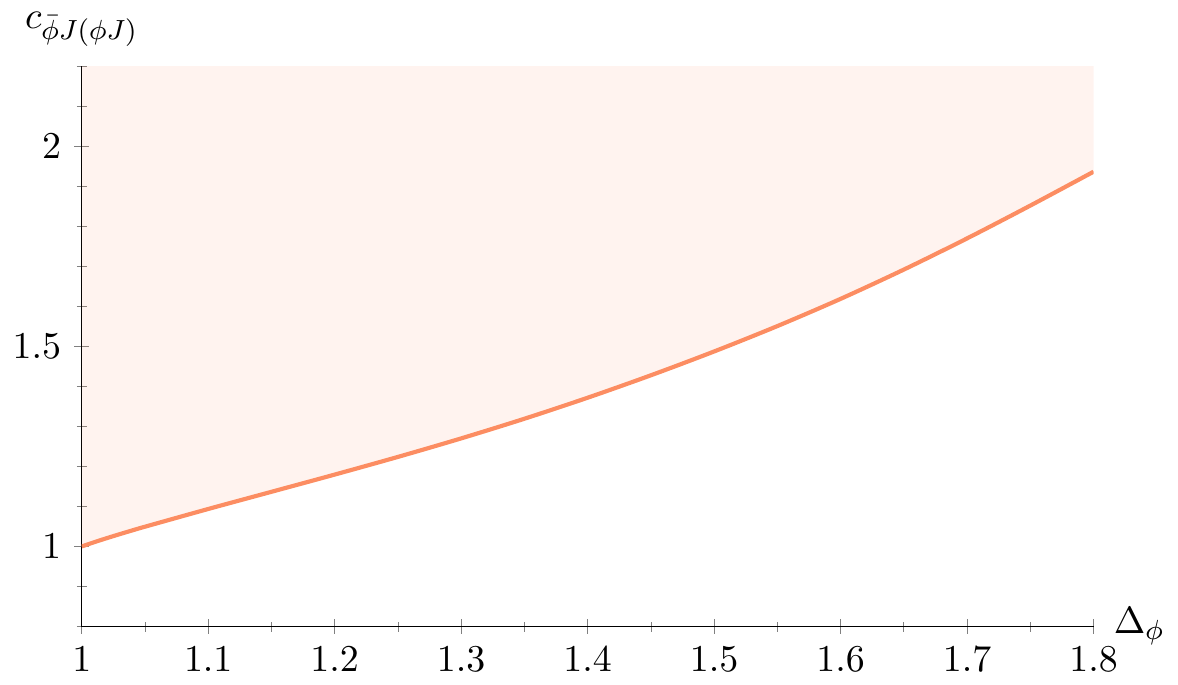}
  \end{center}
  \vspace{-11pt}
  \caption{Upper bound on the OPE coefficient of the operator $\bar{\phi}J$
  in the $\bar{\phi}\times J$ OPE as a function of $\Delta_\phi$, demanding
  $c_{\bar{\phi}J\phi}=c_{\bar{\phi}\phi J}$. In this plot we use
  $\Lambda=17$.}
  \label{fig:phiJ_ope_coeff}
\end{figure}

As we have already seen the second scalar in the $\phib\times J$ OPE has
dimension $\Delta_\phi+2$. We will call it $\phib J$. We can obtain a bound
on its OPE coefficient, again imposing $c_{\phib J\phi}=c_{\phib\phi J}$.
The bound is seen in Fig.~\ref{fig:phiJ_ope_coeff}, and is strongest close
to $\Delta_\phi=1$ where it approaches the expected value of $c_{\phib
J(\phi J)}=1$.

\newsec{Discussion}[Discussion]
This work is the first numerical bootstrap study of mixed correlator
systems in SCFTs with four supercharges. In this paper we focused on 4D
$\mathcal{N}=1$ SCFTs and used the crossing symmetry and positivity in the
$\{\langle \bar{\phi} \phi \bar{\phi} \phi\rangle, \langle\bar{\phi} R \phi
R \rangle, \langle RRRR\rangle\}$ system, where $R$ is a generic real
scalar and $\phi$ is a chiral scalar. We also studied the special case with
$R\to J$, where $J$ is the superconformal primary in a linear multiplet
that contains a conserved global symmetry current. In all these cases we
computed all necessary superconformal blocks, obtaining some new results.

We found new rigorous bounds on 4D $\mathcal{N}=1$ SCFTs that are stronger
than those previously obtained. The features of our results strongly
suggest the existence of a minimal 4D $\mathcal{N}=1$ SCFT with a chiral
operator of dimension $\Delta_\phi\approx1.4$. Nevertheless, further
studies are needed in this system of crossing relations. In particular, we
did not find an isolated island of viable solutions to the crossing
equations similar to that obtained in~\cite{Kos:2014bka, Kos:2015mba}. We
believe that in order to address this more definitively we need to overcome
the current practical limits on the dimension of the functional search
space we can use with the available computational resources. When that
becomes possible, we expect certain dimension bounds to become much more
constraining. However, this will likely require a new level of both
algorithmic efficiency and computational power. We expect to return to this
system when such resource becomes available.

\ack{We would like to thank Zuhair Khandker and David Poland for useful
discussions and collaboration at the initial stages of this project. AS is
grateful to Miguel Paulos, Alessandro Vichi, and Ran Yacoby for useful
discussions. AS also thanks Alessandro Vichi for help with \texttt{SDPB}.
DL thanks Jared Kaplan, Balt van Rees and Junpu Wang for discussions. We
thank the Aspen Center for Physics, supported by the National Science
Foundation under Grant No.~1066293, for hospitality during the initial
stages of this work.  The numerical computations in this paper were run on
the Omega and Grace computing clusters at Yale University, and the LXPLUS
cluster at CERN.  This research is supported in part by the National
Science Foundation under Grant No.~1350180.}

\begin{appendices}

\newsec{Polynomial approximations}[appApprox]
In this work  we consider crossing relations for four-point functions
involving operators with different scaling dimensions $\Delta_1$ and
$\Delta_2$, e.g.\
\eqn{\sum_{\CO}|c|^2\lsp\CF_{\Delta,\lsp\ell,\lsp \Delta_1,\lsp
\Delta_2}(u,v)=0\,,}[genCrossRel]
where $\CF_{\Delta,\lsp\ell,\lsp \Delta_1,\lsp \Delta_2}(u,v)
=u^{-(\Delta_1+\Delta_2)/2}\lsp\CG_{\Delta,\lsp\ell,\lsp \Delta_1,\lsp
\Delta_2}(u,v)-(u\leftrightarrow v)$, with $\CG$ a superconformal block. The
superconformal block contains ordinary conformal blocks defined in
\ConfBlocks. In order to use semidefinite programming techniques we have to
approximate derivatives on $\CF$ and $\CF^{\lsp\prime}$ as positive
functions times polynomials~\cite{Poland:2011ey}. Here we explain how we do
this for expressions like \genCrossRel, assuming first that $\CF$ contains
a single conformal block. To signify this we will use $F$ instead of
$\CF$.\foot{Polynomial approximations of conformal blocks corresponding to
four-point functions involving operators with different scaling dimensions
were recently considered in~\cite{Costa:2016xah}.}

From \ConfBlocks and using $u=z\zb$ and $v=(1-z)(1-\zb)$ we have
\eqn{(z-\zb)F_{\Delta,\lsp\ell,\lsp \Delta_1,\lsp \Delta_2}(z,\zb)=(-1)^\ell
\big(u_{\Delta+\ell}^{\beta,\lsp \gamma,\lsp \delta}(z)\lsp
u_{\Delta-\ell-2}^{\beta,\lsp\gamma,\lsp\delta}(\zb)+
u_{\Delta+\ell}^{\beta,\lsp \gamma,\lsp \delta}(1-z)\lsp
u_{\Delta-\ell-2}^{\beta,\lsp\gamma,\lsp\delta}(1-\zb)\big)
-(z\leftrightarrow \zb)\,,}[Fofu]
where $\beta$ and $\gamma$ can here be either $\Delta_1-\Delta_2$ or
$\Delta_2-\Delta_1$ depending on the four-point function we are
considering, $\delta=\frac12(\Delta_1+\Delta_2)$, and
\eqn{u_{\alpha}^{\beta,\lsp\gamma,\lsp\delta}(z)=
z^{1-\delta}k_{\alpha}^{\beta,\lsp\gamma}(z)\,.}[udefn]
The constants $\alpha,\beta,\gamma,\delta$ have specific relations to
$\Delta,\ell,\Delta_1,\Delta_2$ when appearing in \Fofu, but below we will
keep them general.  As we see the crossing relation \genCrossRel takes a
convenient form in terms of the function
$u_{\alpha}^{\beta,\lsp\gamma,\lsp\delta}(z)$. For our bootstrap analysis
we now need to compute derivatives of
$u_{\alpha}^{\beta,\lsp\gamma,\lsp\delta}$ with respect to $z$ or $\zb$,
and evaluate them at $z=\zb=\frac12$. An easy way to do this is to use a
power series expansion. Indeed, the function
$u_{\alpha}^{\beta,\lsp\gamma,\lsp\delta}(z)$ can be expanded as
\eqn{u_{\alpha}^{\beta,\lsp\gamma,\lsp\delta}(z)=\sum_{n=0}^\infty
C_{\alpha,\lsp\beta,\lsp\gamma,\lsp\delta}^{\lsp n}
\lsp\tfrac{1}{n!}(z-\tfrac12)^n\,,}[uSerExp]
with
\eqn{C_{\alpha,\lsp\beta,\lsp\gamma,\lsp\delta}^{\lsp
n}=2^{n-\frac12\alpha+\delta-1}\frac{\Gamma\big(\tfrac12
(\alpha-2\lsp\delta+4)\big)}{\Gamma\big(\tfrac12(\alpha-2\lsp\delta+4
-2\lsp n)\big)}\lsp{}_3F_2\big(\tfrac12(\alpha-\beta),
\tfrac12(\alpha+\gamma), \tfrac12\alpha-\delta+2\lsp;
\alpha,\tfrac12\alpha-\delta+2-n\lsp;\tfrac12\big)\,.}[coeffsC]

$C_{\alpha,\lsp\beta,\lsp\gamma,\lsp\delta}^{\lsp n}$ as given in \coeffsC
is nonpolynomial and thus not appropriate for our analysis. Hence, we take
an alternate route here, based on that suggested in~\cite{Poland:2010wg}.
Using the hypergeometric differential equation it is easy to verify that
$u_{\alpha}^{\beta,\lsp\gamma,\lsp\delta}$ satisfies the differential
equation
\eqna{\Big(z^2(1-z)\frac{\dtwo}{d z^2}&+\tfrac12
z\big((\beta-\gamma-4\lsp\delta+2)\lsp
z+4\lsp(\delta-1)\big)\frac{d}{d z}\\
&+\tfrac14\big((\beta-2\lsp\delta+2)(\gamma+2\lsp\delta-2)\lsp z
-(\alpha-2\lsp\delta+2)(\alpha+2\lsp\delta-4)\big)
\Big)u_{\alpha}^{\beta,\lsp\gamma,\lsp\delta}(z)=0\,.}[diffeq]
If we use \uSerExp, then taking $n-2$ derivatives on \diffeq and evaluating
at $z=\frac12$ we find the recursion relation
\eqna{C_{\alpha,\lsp\beta,\lsp\gamma,\lsp\delta}^{\lsp n}&=
-(2\lsp n+\beta-\gamma+4\lsp\delta-10)\lsp
C_{\alpha,\lsp\beta,\lsp\gamma,\lsp\delta}^{\lsp n-1}\\
&\quad+\big(4\lsp n\lsp(n-\beta+\gamma-3)\\
&\hspace{1.65cm}+2\lsp \alpha\lsp(\alpha-2)
-\beta\lsp(\gamma+2\lsp\delta-10)+\gamma\lsp(2\lsp\delta-10)-4\lsp\delta
\lsp(\delta-4)-4\big)\lsp
C_{\alpha,\lsp\beta,\lsp\gamma,\lsp\delta}^{\lsp n-2}\\
&\quad+2\lsp(n-2)(2\lsp n-\beta+2\lsp\delta-8)
(2\lsp n+\gamma+2\lsp\delta-8)
\lsp C_{\alpha,\lsp\beta,\lsp\gamma,\lsp\delta}^{\lsp n-3}\,.}[recrel]
This allows us to write
\eqn{C_{\alpha,\lsp\beta,\lsp\gamma,\lsp\delta}^{\lsp n}=
P_n(\alpha,\lsp\beta,\lsp\gamma,\lsp\delta)\lsp 2^{\delta-1}\lsp
k_{\alpha}^{\beta,\lsp\gamma}(\tfrac12)
+Q_n(\alpha,\lsp\beta,\lsp\gamma,\lsp\delta)\lsp 2^{\delta-1}\lsp
(k_{\alpha}^{\beta,\lsp\gamma})'(\tfrac12)\,,}[recrelII]
where $(k_\alpha^{\beta,\lsp\gamma})'$ is the $z$-derivative of
$k_{\alpha}^{\beta,\lsp\gamma}$ and the polynomials $P$ and $Q$ can be
determined from \recrel.

In order to be able to use semidefinite programming we need to further
express appropriately the right-hand side of \recrelII, for it still
involves the nonpolynomial quantities $k_\alpha^{\beta,\lsp\gamma}$ and
$(k_\alpha^{\beta,\lsp\gamma})'$ evaluated at $\frac12$. To proceed, we
perform a series expansion around $z=0$ of
$k_\alpha^{\beta,\lsp\gamma}(\rho)$ and
$(k_\alpha^{\beta,\lsp\gamma})'(\rho)$, where we use the coordinate
$\rho=z/\big(1+\sqrt{1-z}\big)^2$ \cite{Hogervorst:2013sma}.  The expansion
in $\rho$ converges faster than that in $z$. We perform this expansion to a
fixed order $w$ for $k_\alpha^{\beta,\lsp\gamma}$ and $w-1$ for
$(k_\alpha^{\beta,\lsp\gamma})'$, so that both expressions have the same
poles in $\alpha$, and then we substitute $\rho=\rho(\frac12)=3-2\sqrt{2}$.
Then, in the right-hand side of \recrelII we can pull out a positive factor
equal to $\big(2^{-\frac12\alpha}\alpha\lsp
D_w(\alpha)\big)^{-1}$,\foot{Since $\alpha$ is here $\Delta+\ell$ or
$\Delta-\ell-2$ we may have $\alpha=-1$, in which case $\alpha
D(\alpha)=0$. This corresponds to the case where the exchanged operator is
a free scalar.} where $D_w(\alpha)$ is the denominator of the power series
expansion of $k_\alpha^{\beta,\lsp\gamma}$ evaluated at $\rho(\frac12)$.
Doing so we can bring \recrelII to the form
\eqn{C_{\alpha,\lsp\beta,\lsp\gamma,\lsp\delta}^{\lsp
n}\to C_{\alpha,\lsp\beta,\lsp\gamma,\lsp\delta,\lsp w}^{\lsp n}\approx
2^{\delta+\frac12\alpha-1}\frac{1}{\alpha\lsp D_w(\alpha)}\lsp
R_{n,w}(\alpha,\beta,\gamma,\delta)\,,\qquad
\alpha\lsp D_w(\alpha)>0\text{ for }\alpha>-1\,,}[finalC]
where $R_{n,w}(\alpha,\beta,\gamma,\delta)$ is polynomial in its arguments,
given by
\eqn{R_{n,w}(\alpha,\beta,\gamma,\delta)=N_{1,w}(\alpha,\beta,\gamma)\,
P_n(\alpha,\beta,\gamma,\delta) +N_{2,w}(\alpha,\beta,\gamma)\lsp
Q_n(\alpha,\beta,\gamma,\delta)\,,}[]
where $N_{1,w}$ is $2^{-\frac12\alpha}\alpha$ times the numerator of the
power series expansion of $k_\alpha^{\beta,\lsp\gamma}$ evaluated at
$\rho(\frac12)$, and $N_{2,w}$ is the power series expansion of
$(k_\alpha^{\beta,\lsp\gamma})'$ multiplied with
$2^{-\frac12\alpha}\alpha\lsp D_w(\alpha)$. The approximation to
$C_{\alpha,\lsp\beta,\lsp\gamma,\lsp\delta}^{\lsp n}$ in \finalC becomes
better as we increase the order $w$ of the power series expansion of
\recrelII.\foot{In this work we have typically used $w$ around 20.} For the
remainder of this appendix we will ignore the label $w$.

Using \Fofu, \uSerExp and \finalC, derivatives of
$(z-\zb)F_{\Delta,\lsp\ell,\lsp \Delta_1,\lsp \Delta_2} (z,\zb)$ evaluated
at $z=\zb=\frac12$ can now be written as
\eqn{\partial_z^m\partial_{\zb}^n\big((z-\zb)
F_{\Delta,\lsp\ell,\lsp \Delta_1,\lsp
\Delta_2}(z,\zb)\big)\big|_{z=\zb=\frac12}\approx
\chi(\Delta,\ell,\delta)\lsp
U_{m,n}(\Delta,\ell,\beta,\gamma,\delta)\,,}[derApprox]
where
\eqn{\chi(\Delta,\ell,\delta)=\frac{2^{2(\delta-1)+\Delta}}
{(\Delta+\ell)(\Delta-\ell-2)\lsp D(\Delta+\ell)\lsp D(\Delta-\ell-2)}}[]
is positive in unitary theories, and
\eqn{U_{m,n}(\Delta,\ell,\beta,\gamma,\delta)=\tfrac12(1+(-1)^{m+n})(-1)^\ell
\big(R_m(\Delta+\ell,\beta,\gamma,\delta)\lsp
R_n(\Delta-\ell-2,\beta,\gamma,\delta)-(m\leftrightarrow
n)\big)}[UF]
is a polynomial in $\Delta,\ell,\beta,\gamma,\delta$. In the case of $H$
instead of $F$ we find an expression similar to \derApprox but instead of
the overall factor of $1+(-1)^{m+n}$ in \UF we have the factor
$1-(-1)^{m+n}$.

Finally, let us consider derivatives of the function
$\CF_{\Delta,\lsp\ell,\lsp \Delta_1,\lsp \Delta_2}(z,\zb)$ at
$z=\zb=\frac12$. Here we will focus on
$\bar{\CF}_{\Delta,\lsp\ell,\lsp\Delta_\phi-\Delta_R}^{\lsp\phib R\lsp;\lsp
\phi R}(z,\zb)$ of \FphiR, but other $\CF$'s can be treated similarly. We
can again multiply with $z-\zb$ as in \Fofu, and then it is straightforward
to obtain
\eqna{\partial^m_z\partial^n_\zb\big((z-\zb)
\bar{\CF}_{\Delta,\lsp\ell,\lsp\Delta_\phi-\Delta_R}^{\lsp\phib R\lsp;
\lsp \phi R}(z,\zb)\big)\big|_{z=\zb=\frac12}&\approx
\chi(\Delta,\ell,\delta)\\
&\hspace{-4.8cm}\times\big[U_{m,n}(\Delta,\ell,\beta,\gamma,\delta)\\
  &\hspace{-4.3cm}+4\lsp\rho(\tfrac12)\lsp \bar{c}_1
\frac{(\Delta+\ell)\lsp \tilde{D}(\Delta+\ell)}
{(\Delta+\ell+2)\lsp \tilde{D}(\Delta+\ell+2)}\lsp
U_{m,n}(\Delta+1,\ell+1,\beta,\gamma,\delta)\\
&\hspace{-4.3cm}+4\lsp\rho(\tfrac12)\lsp \bar{c}_2
\frac{(\Delta-\ell-2)\lsp \tilde{D}(\Delta-\ell-2)}
{(\Delta-\ell)\lsp \tilde{D}(\Delta-\ell)}\lsp
U_{m,n}(\Delta+1,\ell-1,\beta,\gamma,\delta)\\
&\hspace{-4.3cm}+16\lsp\rho^2(\tfrac12)\lsp \bar{c}_1
\bar{c}_2\lsp
\frac{(\Delta+\ell)(\Delta-\ell-2)\lsp \tilde{D}(\Delta+\ell)
\tilde{D}(\Delta-\ell-2)}{(\Delta+\ell+2)(\Delta-\ell)\lsp
\tilde{D}(\Delta+\ell+2)\lsp \tilde{D}(\Delta-\ell)}
\lsp U_{m,n}(\Delta+2,\ell,\beta,\gamma,\delta)\big]\,,}[superDerApprox]
where $\beta=\gamma=\Delta_\phi-\Delta_R$,
$\delta=\frac12(\Delta_\phi+\Delta_R)$, $\bar{c}_1$ and $\bar{c}_2$ are
given by \cOneTwo, and
$\tilde{D}(\alpha)=\big(2\lsp\rho(\frac12)\big)^{\frac12\alpha}\lsp
D(\alpha)$ is polynomial in $\alpha$. Now, since $\tilde{D}(\alpha)$ is a
polynomial of degree $w$ of the form $\alpha(\alpha+1)\cdots (\alpha+w-1)$,
it is
\eqn{\frac{\alpha\lsp \tilde{D}(\alpha)}
{(\alpha+2)\lsp \tilde{D}(\alpha+2)}=\frac{\alpha^2(\alpha+1)}{(\alpha+2)
(\alpha+w)(\alpha+w+1)}\,.}[]
As a result, \superDerApprox can be written as
\eqna{\partial^m_z\partial^n_\zb\big((z-\zb)
\bar{\CF}_{\Delta,\lsp\ell,\lsp \Delta_\phi-
\Delta_R}^{\lsp\phib R\lsp;\,\phi R}
(z,\zb)\big)\big|_{z=\zb=\frac12}&\approx
\frac{\chi(\Delta,\ell,\delta)}
{f(\Delta+\ell)\lsp f(\Delta-\ell-2)}\\
&\hspace{-1.5cm}\times \big[f(\Delta+\ell)\lsp f(\Delta-\ell-2)\lsp
U_{m,n}(\Delta,\ell,\beta,\gamma,\delta)\\
&\hspace{-1cm}+\bar{c}_1\lsp g(\Delta+\ell)\lsp f(\Delta-\ell-2)\lsp
U_{m,n}(\Delta+1,\ell+1,\beta,\gamma,\delta)\\
&\hspace{-1cm}+\bar{c}_2\lsp f(\Delta+\ell)\lsp g(\Delta-\ell-2)\lsp
U_{m,n}(\Delta+1,\ell-1,\beta,\gamma,\delta)\\
&\hspace{-1cm}+\bar{c}_1\bar{c}_2\lsp g(\Delta+\ell)
\lsp g(\Delta-\ell-2)\lsp U_{m,n}(\Delta+2,\ell,\beta,\gamma,\delta)
\big]\,,}[superDerApproxII]
where
\eqn{f(\alpha)=(\alpha+2)(\alpha+w)(\alpha+w+1)(\alpha+\Delta_\phi)\,,
\qquad
g(\alpha)=4\lsp\rho(\tfrac12)\lsp\alpha^2(\alpha+1)(\alpha+\Delta_\phi)
\,.}[]
The quantity $\chi(\Delta,\ell,\delta)/f(\Delta+\ell)\lsp f(\Delta-\ell-2)$
is positive in unitary theories since $w>1$. Furthermore, the factors in
the denominators of $\bar{c}_1$ and $\bar{c}_2$ are also contained in the
corresponding $g$ that multiplies them in \superDerApproxII. Therefore, the
right-hand side of \superDerApproxII is of the form of a positive quantity
times a polynomial and so it can be used in our bootstrap analysis.

\newsec{On the derivation of superconformal blocks}[appSBlocks]
In this appendix we briefly describe the method we used to compute the
superconformal blocks of section~\ref{secSCB}. Despite significant
developments on $\mathcal{N}=1$ superconformal blocks~\cite{Poland:2010wg,
Fortin:2011nq, Berkooz:2014yda, Fitzpatrick:2014oza, Khandker:2014mpa,
Li:2016chh}, blocks that arise from superdescendants whose corresponding
primaries do not contribute have not been treated systematically. An
example has been worked out in~\cite{Fortin:2011nq}, while, in the case of
interest for this paper, namely regarding the $\bar{\phi} \times R$ OPE, an
example is the superconformal primary $\CO_{\dot{\alpha}}$, which cannot
appear because it does not have integer spin, but whose descendants
$\bar{Q}^{\dot{\alpha}} \CO_{\dot{\alpha}}$ and (the primary component of)
$\bar{Q}^2 Q_{\alpha}\CO_{\dot{\alpha}}$ may both appear and form a
superconformal block.

As mentioned in section \ref{secSCB}, there are two types of such operators
for the four-point function we are interested in. The first has $\jb=j+1$,
that is, it has one more dotted than undotted index.  The superconformal
primary $\CO_{\alpha_1\hspace{-0.8pt}\ldots\alpha_\ell;\lsp\alphad\alphad_1
\hspace{-0.8pt}\ldots\alphad_\ell}$ has zero three-point function with two
scalars because it does not have integer spin. The superdescendant
$Q^{\alpha}\COb_{\alpha\alpha_1\hspace{-0.8pt}\ldots\alpha_\ell;
\lsp\alphad_1\hspace{-0.8pt}\ldots\alphad_\ell}$ has spin $\ell$ and the
primary component of the superdescendant $Q^2 \Qb_{(\alphad}
\COb_{\alpha\alpha_1\hspace{-0.8pt}\ldots\alpha_\ell;\lsp\alphad_1
\hspace{-0.8pt}\ldots\alphad_\ell)}$ has spin $\ell+1$. These two
superdescendants have nonzero three-point function with $\phib$ and $R$ if
the weights of the associated superconformal primary
$\CO_{\alpha_1\hspace{-0.8pt}\ldots\alpha_\ell;\lsp\alphad\alphad_1
\hspace{-0.8pt}\ldots\alphad_\ell}$ satisfy
$q=\frac12(\Delta+\Delta_\phi-\frac32)$ and
$\qb=\frac12(\Delta-\Delta_\phi+\frac32)$.

There is a second class of operators
$\CO_{\alpha_1\hspace{-0.8pt}\ldots\alpha_\ell;\lsp
\alphad_2\ldots\alphad_\ell}$, $\ell\ge1$, that has one more undotted
index. When $q=\frac12(\Delta+\Delta_\phi-\frac32)$ and
$\qb=\frac12(\Delta-\Delta_\phi+\frac32)$, the superdescendant
$Q_{(\alpha_1}\COb_{\alpha_2\ldots\alpha_\ell);\lsp
\alphad_1\hspace{-0.8pt}\ldots\alphad_\ell}$ and the primary component of
$Q^2 \Qb^{\dot{\alpha}}\COb_{\alpha_2\ldots\alpha_\ell;\lsp
\alphad\alphad_2\hspace{-0.8pt}\ldots\alphad_\ell}$ have nontrivial
three-point functions with $\phib$ and $R$.

In this appendix we summarize the calculation of such superconformal blocks
in four-dimensional $\mathcal{N}=1$ SCFTs. We focus on the contribution of
an exchanged superconformal multiplet in the $\bar{\phi} \times R$ channel
of the four-point function $\langle\bar{\phi} R \phi R\rangle$. In $d\ge3$
dimensions, a superconformal multiplet includes a finite number of
conformal multiplets. Therefore, the superconformal block is a linear
combination of conformal blocks with coefficients fixed by supersymmetry.
For each conformal primary component $\CO$ of the supermultiplet, this
coefficient is given by $c_{\bar{\phi} R \CO}c_{\phi R \COb}/c_{\COb\CO} $,
where $c_{\bar{\phi} R \CO}$ and $c_{\phi R \COb}$ are the three-point
function coefficients and $c_{\COb\CO}$ is the two-point function
coefficient. The construction of primary components and their two-point
function coefficients $c_{\COb\CO}$ for any 4D $\mathcal{N}=1$
superconformal multiplet has been worked out in~\cite{Li:2014gpa}. The form
of the superfield three-point function was originally worked out
in~\cite{Park:1997bq, Osborn:1998qu}, and reproduced for the cases of
interest here in~\eqref{ThreePF}, \eqref{tIThetaI} and \eqref{tIThetaII}.
Using the {\it Mathematica} package developed in \cite{Li:2014gpa}, we
expand these three-point functions in $\theta$ and $\bar{\theta}$. Using
the explicit construction of the superfield at each $\theta$,
$\bar{\theta}$ order worked out in \cite{Li:2014gpa}, we match the result
of the expansion of the superfield three-point functions to the expected
form of conformal three-point functions and solve for the three-point
function coefficients $c_{\bar{\phi} R \CO}$.

As an illustration, we elaborate more on this calculation for the first
class of operators mentioned above. Expanding \eqref{ThreePF} with
\eqref{tIThetaI} to first order in $\thetab_3$, we have
\eqn{\vev{\Phib(z_1)\lsp
\CR(z_2)\lsp\CO(z_3,\eta,\bar{\eta})}_{\bar{\theta}_3}=
-i\lsp\frac{1}{r_{13}^{\,2\Delta_{\phi}}r_{23}^{\,2\Delta_{R}}}
(Z_{3}{\hspace{-0.8pt}}^2)^{\frac12(\Delta-\ell+\frac12-\Delta_{\phi}-\Delta_{R})}\left(\eta\rm{Z}_{3}\bar{\eta}\right)^{\ell}\bar{\eta}
\bar{\theta}_{3}\,,}[OITb]
where $r_{ij}=(x_{ij}^{\,2})^{\frac12}$,
$Z_3^\mu=-x_{13}^\mu/x_{13}^{\,2}+x_{23}^\mu/x_{23}^{\,2}$,
$\text{Z}_{3\lsp\alpha\alphad}=Z_{3\lsp\mu}\lsp\sigma^\mu_{\alpha\alphad}$,
$Z_3{\hspace{-0.8pt}}^2=x_{12}^{\,2}/x_{13}^{\,2} x_{23}^{\,2}$, and we
have used bosonic auxiliary spinors $\eta$ and $\bar{\eta}$ to saturate all
free spinor indices on $\CO$:
\eqn{\CO(z,\eta,\bar{\eta})\equiv
\frac{1}{(\ell!)^2}\eta^{\alpha_{1}}\!
\cdots\eta^{\alpha_{\ell}}\bar{\eta}^{\alphad_{1}}\!\cdots
\bar{\eta}^{\alphad_{\ell}}\COind{\alpha}\,.}[COjjb]
Note that the $x$-dependence on the right-hand side of \eqref{OITb} has
exactly the form of a three-point function of conformal primaries. It
corresponds to the contribution from
$\bar{Q}^{\dot{\alpha}}\CO_{\alpha_1\hspace{-0.8pt} \ldots\alpha_\ell;
\lsp\alphad\alphad_1 \hspace{-0.8pt}\ldots\alphad_\ell}$ in the three-point
function. Using the superfield structure worked out in \cite{Li:2014gpa},
\begin{equation}
e^{i\lsp\theta Q+i\lsp\thetab \bar{Q}}\mathcal{O}_{\ell,\lsp\ell+1}
|_{\thetab}=i\thetab\Qb\CO_{\ell,\hspace{1pt}\ell+1}
=i\lsp\thetab\partial_{\etab}(\Qb\CO)_{\ell,\hspace{1pt}\ell+2}
+i\lsp\frac{\ell+1}{\ell+2}\lsp\thetab\etab\lsp
(\Qb\CO)_{\ell,\hspace{1pt}\ell}\,.\label{eq:QbO}
\end{equation}
Here $(\Qb\CO)_{\ell,\hspace{1pt}\ell+2}$ and
$(\Qb\CO)_{\ell,\hspace{1pt}\ell}$ are the two conformal primaries obtained
from symmetrizing or antisymmetrizing the index of $\Qb$ with the dotted
indices of the superconformal primary $\mathcal{O}$. Only the later can
appear in the three-point function with scalars because it has integer
spin. Plugging \eqref{eq:QbO} into the left-hand side of \eqref{OITb} we
find that the three-point function coefficient of $\langle \phib
R(\Qb\CO)_{\ell,\hspace{1pt}\ell}\rangle$ is
\eqn{c_{\phib R(\Qb\CO)}=\frac{\ell+2}{\ell+1}\,.}[lambdaQbO]
To get the three-point function coefficient for the $\Qb^2 Q\CO$
descendant, we first work out the $\theta$-expansion of the superfield
three-point function. The result is
\eqna{\vev{\Phib(z_1)\lsp
\CR(z_2)\lsp\CO_I(z_3,\eta,\bar{\eta})}_{\theta_3\bar{\theta}_3^{\lsp2}}
=\frac{1}{r_{13}^{\,2\Delta_{\phi}}r_{23}^{\,2\Delta_{R}}}&
(Z_{3}{\hspace{-0.8pt}}^{2})^{\frac12(\Delta-\ell+\frac{1}{2}
-\Delta_{\phi}-\Delta_{R})}(\eta\text{Z}_{3}\bar{\eta})^{\ell}\\
&\bar{\theta}_{3}^{\lsp 2}\Big((\Delta_{\phi}-\ell-2)
\frac{1}{x_{13}^{\,2}}\theta_{3}\text{x}_{13}\bar{\eta}\\
&\qquad-\tfrac{1}{4}\big(2(\Delta+3\ell-\Delta_\phi-\Delta_R)+9\big)
\theta_{3}\text{Z}_{3}\bar{\eta}\Big)\,.
}[]
This does not take the form of a three-point function involving conformal
primaries. This is expected since at this order in $\theta$ and $\thetab$
the three-point function also contains contributions from conformal
descendants. In particular, following notation of~\cite{Li:2014gpa}, we
have
\eqna{e^{i\theta Q + i\thetab
\Qb}&\mathcal{O}_{\ell,\hspace{1pt}\ell+1}\big|_{\theta\bar{\theta}^2}
=-\frac{i}{4}\thetab^{\hspace{0.5pt}2}\,\theta\partial_{\eta}\left((\Qb^{2}
Q\CO)^\eta_{\ell+1,\hspace{1pt}\ell+1;\hspace{1pt}p}+
2i\bar{c}_{5}\,\partial_{\bar{\eta}}\partial_{x}\eta\,(\Qb\CO)_{\ell,\hspace{1pt}\ell+2\hspace{1pt};\hspace{1pt}p}\right.\\
&\hspace{8cm}\left.-2i\bar{c}_{6}\lsp\frac{\ell+1}{\ell+2}\lsp\eta
\partial_{x}\bar{\eta}\,(\Qb\CO)_{\ell,\hspace{1pt}\ell;\hspace{1pt}p}
\right)\\
&\quad\hspace{-0.5cm}+\frac{i}{4}\frac{\ell+1}{\ell+2}
\thetab^{\hspace{0.5pt}2}\,
\theta\eta\left((\Qb^{2}Q\CO)_{\ell-1,\hspace{1pt}\ell+1;\hspace{1pt}p}+
2i\bar{c}_{7}\,\partial_{\eta}\partial_{x}\partial_{\bar{\eta}}\,
(\Qb\CO)_{\ell,\hspace{1pt}\ell+2;\hspace{1pt}p}-
2i\bar{c}_8\lsp\frac{\ell+1}{\ell+2}\,\etab\partial_{x}\partial_{\eta}\,(\Qb\CO)_{\ell,\hspace{1pt}\ell;\hspace{1pt}p}\right),}[eq:Q2QbO]
and we see that two different descendants have integer spins and can
contribute to the three-point function with $\phib$ and $R$. The relevant coefficients can be obtained from \cite{Li:2014gpa}:
\eqn{\bar{c}_6=-2\frac{\Delta_\phi-2}{(\ell+1)\big(2(\Delta+\ell)+1\big)}
\,,
\qquad \bar{c}_8=-2\frac{\Delta_\phi-\ell-3}{\ell\lsp(2\Delta-1)}\,.}[]
Removing these contributions from the superfield correlator we indeed get
a conformal primary three-point function with coefficient
\eqn{c_{\phib R(\bar{Q}^2 Q\CO)_p}=
i\lsp\frac{\big(2(\Delta-\Delta_\phi+\ell)+5\big)
\big(2(\Delta+\Delta_\phi-\Delta_R+\ell)+1\big)}
{(\ell+1)\big(2(\Delta+\ell)+1\big)}\,.}[]
Finally, using the two-point function coefficient derived in
\cite{Li:2014gpa}, we get the results \eqref{blockThetaI} and \eqref{chat}.
For the second class of operators we carried out a similar procedure and
obtained \eqref{blockThetaII} and \eqref{ccheck}.

Although we will not present the details here, this calculation is easily
generalized to cases where the operator $R$ is not real and carries an
R-charge. The relevant results can be found in \eqref{hatcS} and
\eqref{checkcS}.  More generally, for other scalar $\mathcal{N}=1$
superconformal four-point functions, there may be intermediate operators of
this type that do not correspond to \eqref{tIThetaI} or \eqref{tIThetaII}.
We have not calculated such superconformal blocks, but our method should
apply straightforwardly to such cases. Indeed, this method is a feasible
way of computing any $\mathcal{N}=1$ scalar superconformal block in a case
by case basis.

\end{appendices}

\bibliography{BootstrapN1SCFTs}

\end{document}